\documentclass[%
reprint,
superscriptaddress,
frontmatterverbose,
showpacs,preprintnumbers,
 amsmath,amssymb,
 aps,pra,Ifloatfix,
citeautoscript,
]{revtex4-2}

\usepackage{mdframed}

\usepackage[pdftex]{graphicx}

\usepackage{mathrsfs}
\usepackage[linkcolor=blue,colorlinks=true,citecolor=blue,filecolor=blue]{hyperref}
\usepackage{amsmath}
\usepackage[english]{babel}
\usepackage{booktabs}
\usepackage{amssymb}
\usepackage{type1cm}
\usepackage{braket}
\usepackage{tikz}
\usepackage{pgfplots}
\usepackage{lipsum}
\usepackage{xcolor}
\usepackage{graphicx}% Include figure files
\usepackage{subcaption}
\usepackage{dcolumn}% Align table columns on decimal point
\usepackage{bm}% bold mathd
\usepackage{physics}
\usepackage[utf8]{inputenc}
\usepackage{multirow}
\usepackage{tikz}
\definecolor{Midnight_Blue}{rgb}{0.1, 0.1, 0.6}

\usepackage{cleveref}
\crefname{equation}{Eq.}{Eqs.}
\Crefname{equation}{Equation}{Equations}
\crefname{figure}{Fig.}{Figs.}
\Crefname{figure}{Figure}{Figures}
\crefname{figure}{Fig.}{Figs.}
\Crefname{figure}{Figure}{Figures}
\crefname{section}{Supplemental Material Section}{Supplemental Material Sections}
\Crefname{section}{Supplemental Material Section}{Supplemental Material Sections}
\crefname{appendix}{Appendix}{Appendices}
\Crefname{appendix}{Appendix}{Appendices}
\crefname{table}{Table}{Tables}
\Crefname{table}{Table}{Tables}

\usepackage{enumitem,amssymb}
\newlist{todolist}{itemize}{2}
\setlist[todolist]{label=$\square$}
\usepackage{pifont}

\usepackage{etoolbox}
\pgfplotsset{compat=1.18} 
\begin{document}

\title{Loss-Tolerant Quantum Communication via Bosonic-GKP-Parity-Encoding}

\author{S. Nibedita Swain}
\email{swain.snibedita@gmail.com}
\affiliation{School of Mathematical and Physical Sciences,
University of Technology Sydney, Ultimo, NSW 2007, Australia}

\affiliation{Sydney Quantum Academy, Sydney, NSW 2000, Australia}
\affiliation{Centre for Quantum Computation and Communication Technology, School of Mathematics
and Physics, University of Queensland, Brisbane, Queensland 4072, Australia}

\author{Timothy C. Ralph}
\email{ralph@physics.uq.edu.au}
\affiliation{Centre for Quantum Computation and Communication Technology, School of Mathematics
and Physics, University of Queensland, Brisbane, Queensland 4072, Australia}

\begin{abstract}
Quantum repeaters constitute a promising platform for enabling long-distance quantum communication 
and may ultimately serve as the backbone of a secure quantum internet, a scalable quantum network, 
or a distributed quantum computer. An efficient approach to encoding qubits within an error-
correcting code is provided by bosonic codes, in which even a single oscillator mode can function 
as a sufficiently large physical system. In this work, we initially investigate the bosonic 
Gottesman–Kitaev–Preskill (GKP) code as a promising platform for loss correcting quantum 
repeaters, compatible with room temperature implementation, and analyse how loss and other noise 
sources propagate through the circuit using Heisenberg evolution.
We analyse three quantum repeater protocols in which transmission loss is suppressed at the cost 
of logical errors, identifying a relay-like teleamplifier as the optimal scheme. This enables long-
distance quantum communication via densely packed nodes without higher-level encoding, and we 
evaluate the resulting secure key rates exploiting analog syndrome information.
Furthermore, we propose a concatenated Bell-state measurement (CBSM) scheme with a modified parity 
encoding based on GKP qubits, CV measurement with teleamplifier and a clipping method  that 
corrects transmission loss without introducing logical errors. This significantly enhances the 
possible secure key distance. We find that GKP-based repeaters can achieve performance comparable 
to approaches relying on photonic qubits, while requiring orders of magnitude fewer qubits. Our 
parity-encoded GKP repeater protocol achieves a substantially higher secret key rate than existing 
GKP-based repeater schemes while maintaining a comparable resource overhead.

\end{abstract}

\date{\today}

\frenchspacing

\maketitle

\section{Introduction} 
Reliable quantum communication protocols are a fundamental prerequisite for realizing a secure quantum internet \cite{kimble2008quantum, wehner2018quantum, azuma2023quantum}, as well as for implementing secure classical communication and distributed quantum cryptographic protocols \cite{scarani2009security, %weedbrook2010quantum, 
pirandola2020advances}. At present, medium-distance quantum communication is routinely demonstrated using qubits encoded with single photons \cite{ma2012quantum, bhaskar2020experimental, slussarenko2022quantum, zhan2025experimental}. However, such photonic qubits are inherently susceptible to complete loss into the environment, photon loss, with the probability of successful transmission (often referred to as the channel efficiency) decaying exponentially with transmission distance. Quantum repeaters get the better of this unfavourable scaling by subdividing long communication channels into shorter, more manageable segments. When stationed between two parties, conventionally denoted as the sender Alice and the receiver Bob, quantum repeater protocols replace the exponential decay of transmission efficiency with a polynomial scaling with the total separation distance \cite{yamamoto1994quantum, briegel1998interdisciplinary, van2006hybrid, dias2017quantum, li2019experimental}. The highest predicted data rates occur for third generation or one-way quantum repeaters for which all the quantum and classical data flows in one direction \cite{munro2012quantum, azuma2015all, borregaard2020one, lee2021loss}.

In recent years, there has been significant interest in bosonic quantum error correcting schemes, which encode a finite-dimensional quantum system into a harmonic oscillator, such as cat codes and Gottesman–Kitaev–Preskill (GKP) codes \cite{cochrane1999macroscopically, PhysRevA.64.012310, lund2008fault}. This renewed attention has been driven by experimental demonstrations of their first practical implementations \cite{ofek2016extending, fluhmann2019encoding, konno2024logical}, which in some cases already outperform simple encoding strategies.

The GKP encoding is particularly well suited for protecting quantum information against Gaussian random displacement noise \cite{PhysRevA.64.012310, fukui2017analog, tzitrin2020progress}. This property can be extended to manage photon loss by transmitting the GKP state through an amplifier, either preceding or following the loss \cite{albert2018performance}. The joint action of the loss and amplification channels effectively results in an unknown Gaussian random displacement, which can be corrected using the standard GKP error correction procedure \cite{PhysRevA.64.012310, glancy2006error}. As a result GKP codes have also emerged as a compelling framework for quantum communication, as they can be encoded in electromagnetic light fields, which are the ideal carriers for long-distance quantum information with the channel loss being mitigated by the GKP error correction \cite{fukui2021all, rozpkedek2021quantum, schmidt2024error, haussler2025quantum}. Moreover, quantum communication protocols require only Clifford operations and Pauli measurements, both of which can be implemented within the GKP encoding using only Gaussian optics and homodyne detection. %Finally, a recent work introduced a method for analysing bosonic quantum circuits in the Heisenberg picture, enabling a useful decomposition of the system’s evolution into signal and noise contributions. This approach provides clear insight into how loss and noise propagate through the circuit \cite{ralph2024noise}.

Many of these schemes can be couched as local teleamplification protocols whereby the incoming state is teleported onto a ``fresh" GKP qubit within the repeater station, removing the loss in the process. Recently, in the context of Gaussian state quantum communication, it has been shown that conceptually similar teleamplification schemes can be enhanced by distributing the teleamplification such that the measurement part occurs at a relay station half way along the channel segment, with the state being teleported to the subsequent repeater station with the help of a classical channel \cite{winnel2021, guanzon2022ideal, swain2025enhancing}. This results in an effective doubling of the distance between repeater stations for the same communication rate.

In the first part of this work, we introduce a number of GKP based teleamplification, one-way repeater protocols, including ones using distributed teleamplification. We find that protocols using the distributed relay scheme perform best and (using secret key generation rates as a metric) can in principle allow quantum communication over distances of a few hundred kilometres.

GKP error correction against loss inevitably introduces some level of logical qubit errors. These errors build up along the channel and ultimately limit the distance quantum information can be communicated. In order to correct the logical errors and hence extend the range of quantum communication higher level encodings are required. 
%Photons are ideal carriers of quantum information; however, photon loss remains the primary limitation for long-distance quantum communication. Since any form of amplification necessarily introduces additional noise into the system \cite{caves1982quantum}, amplification-based strategies for dealing with photon loss are generally not optimal. 
To address this challenge, Bell-state measurement (BSM) based schemes have been proposed. These can be thought of as teleamplification schemes with higher level encoding. When incorporated into quantum repeater architectures they enable communication ranges that scale polynomially with distance for various types of qubits, including single photon and coherent state qubits \cite{lee2018fundamental, lee2021loss}. In the context of quantum computing Fukui \textit{et al.} \cite{fukui2023efficient} proposed concatenating a GKP code with a quantum parity code to identify and discard error prone qubits, together with a linear-time decoding algorithm to correct the logical errors. 

In the second part of this work we introduce a GKP parity encoding scheme designed to mitigate photon loss without introducing logical errors. One approach would be to directly adapt the existing parity-encoding approaches to GKP whereby discrete variable (DV) Bell measurements are employed using photon-number resolving detectors (PNRDs). We demonstrate that this is not the optimal strategy for GKP states. Instead, we employ the strategy of Fukui \textit{et al.} \cite{fukui2023efficient} of implementing continuous-variable (CV) Bell measurements, useing homodyne detection and utilizing the analog information produced. We find this is significantly more effective. Furthermore, this approach is experimentally more accessible and easier to implement than the PNRD-based alternative.
%Collectively, we propose a three-stage quantum repeater protocol based on GKP-encoded qubits. As in any repeater architecture, the transmission distance between Alice and Bob is divided into shorter segments, with repeater stations located at the endpoints of each segment. Quantum information is conveyed from Alice to Bob via sequential teleportation between neighbouring stations, supplemented by classical communication. 
Based on this GKP parity encoding we develop an encoded one-way repeater protocol, in which both loss and logical errors are corrected, enabling quantum communications distances of thousands of kilometres.
%quantum states and classical information propagate unidirectionally toward Bob, from one repeater station to the next. The GKP error-correction procedure, implemented as part of the teleportation step in our repeater protocol, naturally generates analog syndrome information, which can be exploited to enhance the decoding performance of the underlying physical qubits \cite{fukui2017analog}. Finally, we introduce a GKP-parity-encoding scheme, a loss-tolerant code designed to protect GKP qubits against transmission loss.

The remainder of this paper is organized as follows. In Sec. II, we present our teleamplication quantum repeater protocol, which is structured into three related protocols. In Sec. III, we analyse quantum repeater protocols incorporating parity encoding and provide numerical evaluations of the associated resource costs. Finally, Sec. IV is devoted to discussion and conclusions.

\section{Quantum Repeater Scheme with GKP qubits}

\begin{figure*}[!htb]
	\centering
		\includegraphics[scale=0.6]{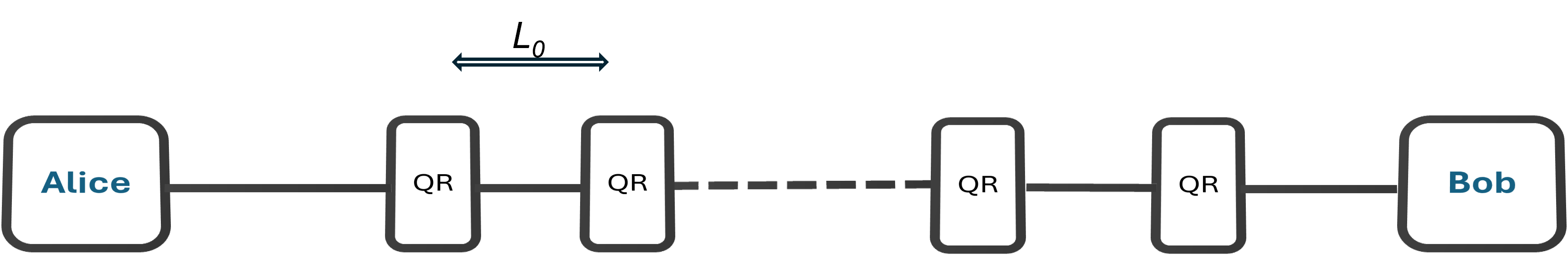}
\caption{ 
A schematic illustration of the one-way quantum repeater protocol. The sender, Alice, prepares a GKP qubit and transmits it to the first quantum repeater station (QR). This station applies protocol-I or protocol-II %, protocol-III) of the protocol 
before forwarding the state to the next station. Each subsequent quantum repeater station repeats the process. %sequentially receives the state, performs protocol-I (protocol-II, protocol-III), and then sends the state onward. 
After passing through the MQR-th quantum repeater station, the qubit is finally received by Bob. For protocol-III the measurement (or relay) stations are distributed half-way between the QR stations which contain the resource states. 
\label{fig:mainQR}}

\end{figure*}
\begin{figure*}[!htb]
	\centering
		\includegraphics[scale=1.5]{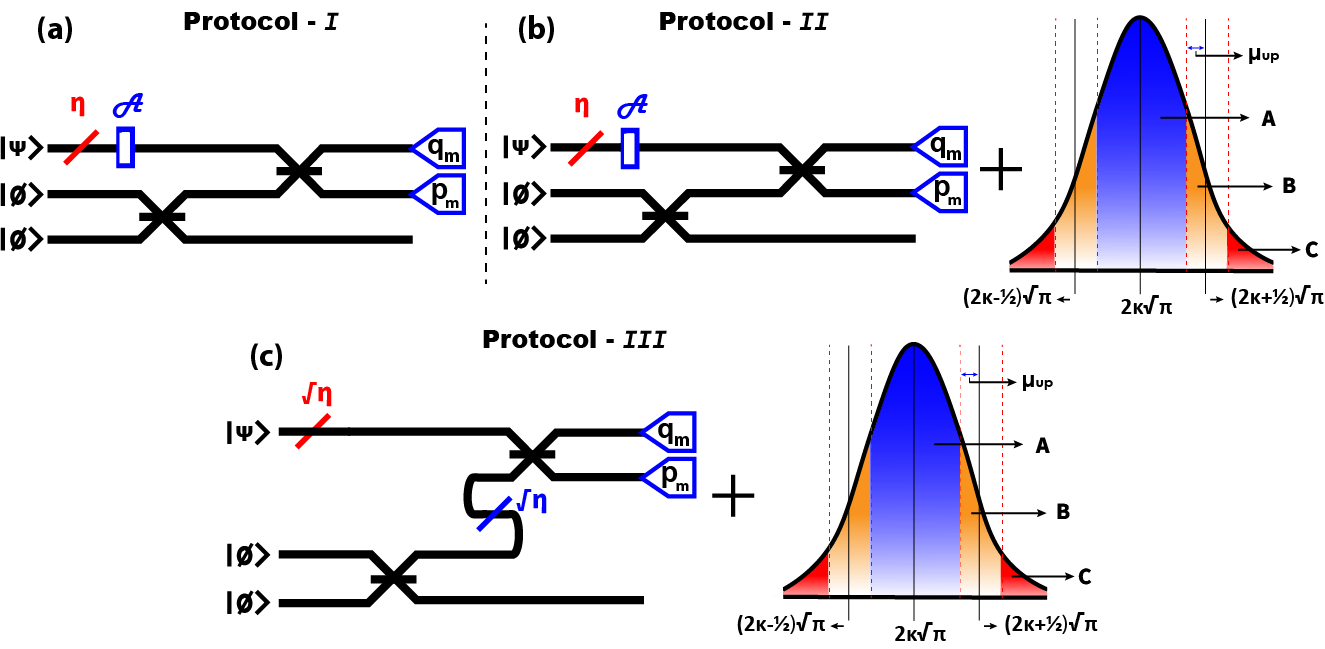}
\caption{ 
A schematic illustration of protocol-I, protocol-II, and protocol-III of our QR protocol. The transmission between QR stations is $\eta$, $A$ are amplifiers and $q_m$ and $p_m$ are position and momentum homodyne measurements respectively.
(a) depicts protocol-I, in which each QR performs amplification followed by imperfect teleportation-based GEC.% together with amplification. 
(b) shows protocol-II, which augments protocol-I by introducing an additional clipping procedure where Region A corresponds to correct bit detection; Region B is where the measurement outcome is flagged as unreliable; and
Region C corresponds to incorrect bit detection. This extension increases the achievable transmission distance at the cost of a reduced success probability. Finally, (c) illustrates protocol-III, which incorporates a distributed set-up with the measurement (or relay) stations half-way between the repeater stations. The resource state to be measured counter-propagates back to the measurement station. %relay-like amplification applied to one of the resource states during the GEC process, 
The effect when combined with the clipping procedure is to enhance the overall communication rate.
\label{fig:Stg123}}
\end{figure*}

We present our quantum repeater (QR) protocols based on GKP qubits in a sequence of protocols, with the aim of providing a clear and intuitive description for the reader. In any repeater protocol, the transmission link between Alice and Bob is divided into shorter segments, with quantum repeater stations placed at the endpoints of each segment. In GKP-based repeater protocols, GKP encoded entangled states may be distributed between adjacent repeater stations via quantum transmission. Information is then conveyed from Alice to Bob through sequential teleportation from one station to the next, assisted by classical communication. In this work, we focus on a one-way repeater protocol in which both quantum states and classical information propagate unidirectionally toward Bob, passing successively from one repeater station to the next, albeit with counter propagating states between repeater and relay nodes in the distributed protocol. We have discussed GKP qubits in Appendix \ref{Gottesman-Kitaev-Preskill qubit} and teleportation-based GKP error correction in Appendix \ref{Teleportation-based GKP Error Correction}. We also note that protocol I and II correspond with the one-way post amplified protocols in Ref.\cite{fukui2021all}.

\subsection{protocol-I}
Fig.\ref{fig:Stg123}(a) presents a schematic of the protocol-I repeater protocol employed in this study. Initially, Alice prepares a GKP qubit and sends it through a photon-loss channel, which represents the dominant source of noise with efficiency $\eta$, to the initial quantum repeater. This process can be modelled as an unintended beamsplitter interaction with an environmental vacuum state. Following propagation through the loss channel, the variances in the $\hat{q}$ and $\hat{p}$ quadratures evolve according to
\begin{align}
    \sigma^{2}_{q} & \to \eta \sigma^{2}_{q} + \frac{1- \eta}{2},\\
    \sigma^{2}_{p} & \to \eta \sigma^{2}_{p} + \frac{1- \eta}{2}
\end{align}
where we use the convention that $\hbar = 1$ (i.e. the variance of the vacuum is $1/2$). Here, $\sigma^{2}_{q}$ and $\sigma^{2}_{p}$ denote the variances of the GKP qubits prior to transmission through the loss channel. In this work, we assume that $\sigma^{2}_{q}$ and $\sigma^{2}_{p}$ are equal, given by $\frac{1}{2}e^{-2r}$, where $r$ is the squeezing parameter. The channel efficiency $\eta$ is related to the transmission distance $L$ (in km) as $\eta = \exp(-L/L_{\text{att}})$, with an attenuation length $L_{\text{att}} = 22$ km.
Photon loss affects the quadrature values in two ways: it attenuates their amplitudes, pulling them toward the origin, and it adds Gaussian noise. Linear amplification provides a convenient method to restore the mean values of the quadratures. An amplification channel can simultaneously restore both quadratures, but at the cost of introducing additional Gaussian noise. A photon-loss channel followed by an amplifier, applied before the teleportation-based GKP error correction (GEC), is effectively equivalent to a Gaussian noise channel with variance $\frac{1-\eta}{\eta}$.
Next, the first repeater receives the GKP qubit and applies the teleportation-based GEC protocol (see Appendices \ref{Gottesman-Kitaev-Preskill qubit}, \ref{Teleportation-based GKP Error Correction}, \ref{Teleportation-based GKP Error Correction with loss} and \ref{Teleportation-based GKP Error Correction with telemplification} for more detail). This procedure corrects deviations arising both from  the finite squeezing of the initial GKP qubit and from the noise introduced by photon amplification by measuring position and momentum mod $\sqrt{\pi}$. After the teleportation-based GEC, the first repeater forwards the qubit to the second repeater. In subsequent move, the $j^{\text{th}}$ repeater ($j = 2, 3, ..., \text{MQR}$) receives the qubit and, after applying the teleportation-based GEC protocol, sends it to the $(j+1)^{\text{th}}$ repeater (as in Fig.\ref{fig:mainQR}), following the same procedure as in previous step. Finally, Bob receives the qubit, optionally performs the teleportation-based GEC protocol, and then measures it.

The linear-optics implementation of GKP error correction, illustrated in Fig.\ref{fig:Stg123}(a) and detailed in Appendix \ref{Teleportation-based GKP Error Correction}, allows us to understand its action on the input state. The probability of misidentifying the bit value of an approximate GKP qubit, denoted by $E_{X (Z)}(\sigma^2)$, is approximately given by $E_{X (Z)}(\sigma^2) = 1-  \int^{\sqrt{\pi}}_{-\sqrt{\pi}} \,dx \frac{1}{\sqrt{2 \pi \sigma^{2} }} e^{-x^2/ 2\sigma^{2}} $. 
While performing GEC, it introduces additional Gaussian noise due to the finitely squeezed ancilla states, the resulting logical error rate for protocol-I (see Fig.~\ref{fig:Stg123}(a)) becomes $E_{X (Z)}(2\sigma^{2} + \frac{1- \eta}{\eta})$. Afterward, the variances of the GKP states are reset to $(\sigma^{2}_{q},\: \sigma^{2}_{p} ) = (\sigma^{2},\: \sigma^{2})$.

\subsection{Protocol-II}
\label{protocol-II}

As pointed out in Ref. \cite{fukui2017analog}, the probability of a logical error in GKP error correction depends on the continuous values of each syndrome measurement, rather than solely on the binned outcome. In other words, syndrome measurements that fall closer to the midpoints between the peaks of the GKP state are less reliable. Discarding GKP qubits that yield such untrustworthy syndromes improves the average quality of the remaining qubits. This principle underlies the clipping method.
During the measurement of a GKP qubit, one determines the bit value $n(=0,1)$ from the measured outcome $q_m = q_n + \delta_m$. In the conventional scheme, the maximum allowable deviation is set as $|\delta_m| \leq \sqrt{\pi}/2$, and the bit value is assigned as $i = (2t + n)\sqrt{\pi}$ ($t = 0, \pm1, \pm2, \dots$). This decision is correct as long as the amplitude of the true deviation $|\bar{\delta}|$ lies between $0$ and $\sqrt{\pi}/2$.
In this method, an upper limit $\mu_{\text{up}} , (< \sqrt{\pi}/2)$ is set to define the maximum deviation that will still yield a correct measurement. If the condition $|\delta_m| < \mu_{\text{up}}$ is violated, the result is discarded. Since measurement errors occur when $|\bar{\delta}| > \sqrt{\pi}/2 + \mu_{\text{up}}$, increasing $\mu_{\text{up}}$ reduces the error probability, albeit at the cost of a lower success probability.
We denote $P^{\text{C}}_{\text{Clip}}$ and $P^{\text{INC}}_{\text{Clip}}$ as the probabilities of a correct and incorrect measurement, respectively, for a GKP qubit with variance $\sigma^{2}$ as in \cite{fukui2017analog}.

\begin{align}
    P^{\text{C}}_{\text{Clip}} & = \sum^{\infty}_{n = -\infty} \int^{\frac{(4n + 1)}{2}\sqrt{\pi} - \mu_{\text{up}}}_{\frac{(4n -1)}{2}\sqrt{\pi} + \mu_{\text{up}} } \,dx \frac{1}{\sqrt{2 \pi \sigma^{2} }} e^{-x^2/ 2\sigma^{2}} \\
 P^{\text{INC}}_{\text{Clip}} & = \sum^{\infty}_{n = -\infty} \int^{\frac{(4n + 3)}{2}\sqrt{\pi} - \mu_{\text{up}}}_{\frac{(4n + 1)}{2}\sqrt{\pi} + \mu_{\text{up}} } \,dx \frac{1}{\sqrt{2 \pi \sigma^{2} }} e^{-x^2/ 2\sigma^{2}}
\end{align}
The probability $P_{\mu_{\text{up}}}$ to obtain the correct bit value with the clipping is given by
\begin{align}
    P_{\mu_{\text{up}}} = \frac{P^{\text{C}}_{\text{Clip}}}{P^{\text{C}}_{\text{Clip}} + P^{\text{INC}}_{\text{Clip}} }
\end{align}
The probability of misidentifying the bit value with the clipping, $E_{\mu_{\text{up}}}$, is given by
\begin{align}
    E_{\mu_{\text{up}}} = 1-   P_{\mu_{\text{up}}}
\end{align}
Then, the success probability of the clipping, $P^{\text{Suc}}_{\mu_{\text{up}}}$, is given by
\begin{align}
    P^{\text{Suc}}_{\mu_{\text{up}}} =  P^{\text{C}}_{\text{Clip}} +  P^{\text{INC}}_{\text{Clip}}
\end{align}

For protocol-II (Fig.\ref{fig:Stg123}(b)), the procedure follows that of protocol-I. However, in the second part of protocol-I, we implement the teleportation-based GEC protocol with the clipping method described above. Additionally, the quantum communication attempt is aborted if the measurement outcome exceeds $\mu_{\text{up}}$. For Protocol-II, in which the clipping method is employed, the resulting logical error rates are given by $E^{X (Z)}_{\mu_{\text{up}}}(2\sigma^2 + \tfrac{1-\eta}{\eta})$. Afterward, the variances of the GKP states are reset to $(\sigma^{2}_{q},\: \sigma^{2}_{p} ) = (\sigma^{2},\: \sigma^{2})$.

\subsection{Protocol-III}
Figure \ref{fig:Stg123}(c) illustrates the schematic of protocol-III. The procedure in this protocol follows a similar sequence as in protocols I and II. However, 
now the GKP error-correction protocol is distributed along the channel with the measurements situated half-way between the repeater stations. 
As a result, both the modes transmitted to the measurement station experience the same loss while being propagated (counter-propagated) over only half of the link distance, $L_0/2$. Consequently, each loss channel is characterised by an effective transmission efficiency of $\sqrt{\eta}$ and the need for amplification is removed because the two modes entering the measurement station are already balanced. 
For protocol–III, when imperfect teleportation-based GEC is implemented, the corresponding logical error rate is given by $E^{X (Z)}_{\mu_{\text{up}}}(2\sigma^{2} +  \frac{(1 - \sqrt{\eta} )}{\sqrt{\eta}} )$. Afterward, the variances of the GKP states are reset to $(\sigma^{2}_{q},\: \sigma^{2}_{p} ) = (\sigma^{2},\: \sigma^{2})$.

\begin{figure*}[!htb]
	\centering
		\includegraphics[scale=2.0]{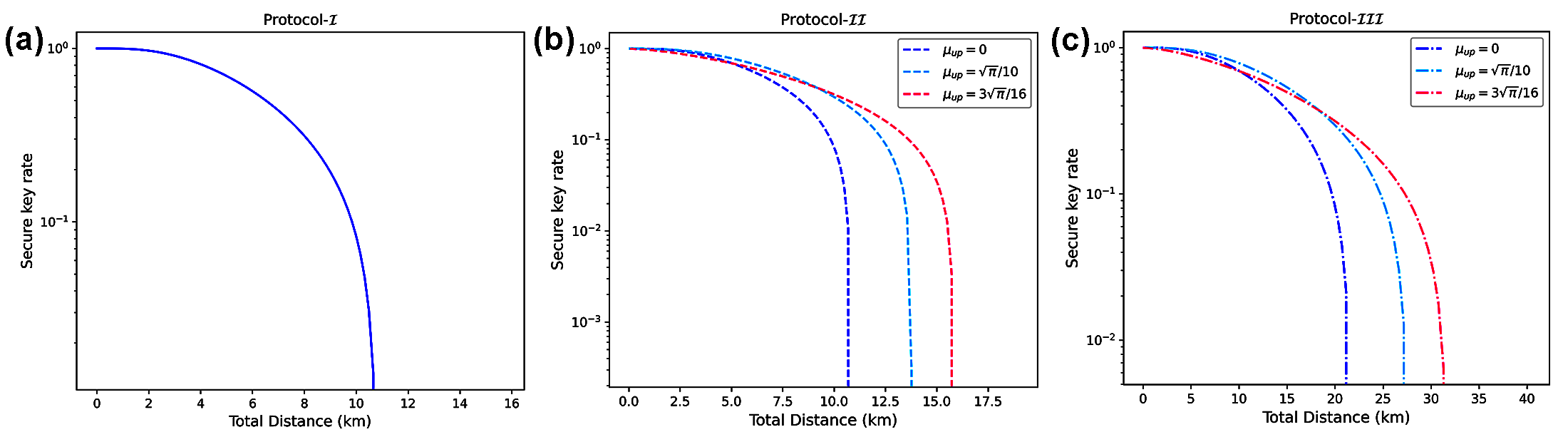}
\caption{ The secure key rate plotted for the one-way quantum repeater protocol in three different protocols for the number of repeater, MQR = 1. $\mu_{\text{up}}$ is the clipping threshold.
\label{fig:s1}}
\end{figure*}

\begin{figure*}[!htb]
	\centering
		\includegraphics[scale=2.0]{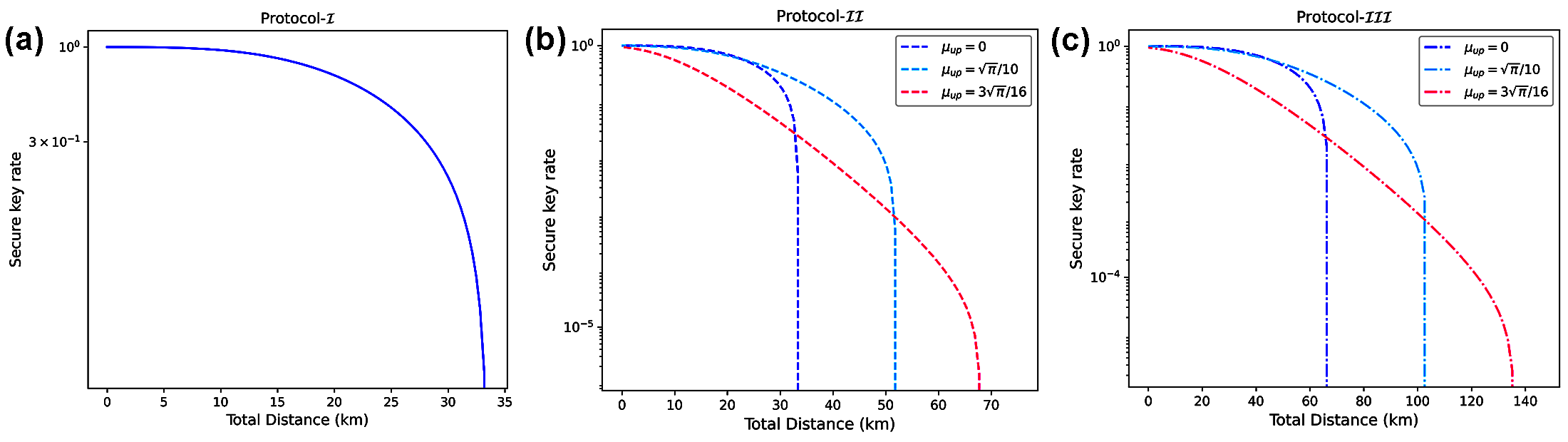}
\caption{The secure key rate plotted for the one-way quantum repeater protocol in three different protocols for the number of repeaters, MQR = 20.
\label{fig:s20}}
\end{figure*}

\begin{figure*}[!htb]
	\centering
		\includegraphics[scale=2.0]{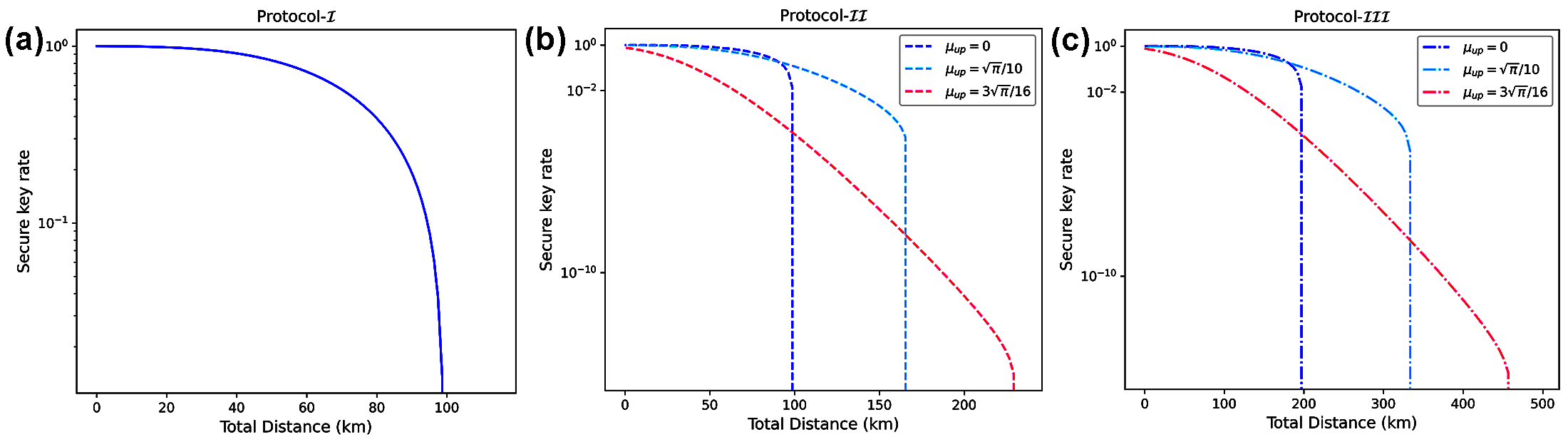}
\caption{The secure key rate plotted for the one-way quantum repeater protocol in three different protocols for the number of repeaters, MQR = 100.
\label{fig:s100}}
\end{figure*}

\subsection{Security Analysis }

Here, we use the logical $X$- and $Z$-error probabilities resulting from teleportation-based GEC performed at each repeater station to evaluate the secret key rates of the repeater protocols considered in this work. These logical $X$ and $Z$ errors are assumed to occur independently and with equal probability. Let $E_X$ and $E_Z$ denote the logical $X$ and $Z$ error probabilities introduced at a single repeater station. The accumulated logical error probabilities over the entire transmission from Alice to Bob are then denoted by $E^{AB}_X$ and $E^{AB}_Z$ for $X$ and $Z$ errors, respectively.
The secure key rate $\kappa \geqq 0$ is given by \cite{scarani2009security, muralidharan2014ultrafast}
$\kappa =   P^{\text{Suc}}_{\mu_{\text{up}}}\{1 - h(E^{AB}_{X}) - h(E^{AB}_{Z})\}$
where $h(\cdot)$ is the binary entropy function and $P^{\text{Suc}}_{\mu_{\text{up}}}$ denotes the success probability of the protocol. In all cases considered here, we have $E_X = E_Z$, and consequently the accumulated error probabilities can be approximated as \cite{muralidharan2014ultrafast}
$E^{AB}_{X} = E^{AB}_{Z} \equiv \frac{1}{2} \{1 - (1 - 2E_{X})^{\text{MQR}} \}$.
When the clipping method is not employed, the success probability satisfies $P^{\text{Suc}}_{\mu_{\text{up}}} = 1$. In contrast, applying the clipping method yields $P^{\text{Suc}}_{\mu_{\text{up}}} < 1$, but can nevertheless enhance the secure key rate by reducing the logical error probabilities $E_X$ and $E_Z$. We assume an initial GKP squeezing level of $15.0$~dB, and model the channel transmissivity $\eta$ as a function of the transmission distance $L$ (in km) according to $\eta = \exp(-L/L_{\text{att}})$, with an attenuation length of $L_{\text{att}} = 22$~km.
In our numerical simulations, we consider the number of quantum repeater stations to be $\text{MQR} = 1, 20,$ and $100$, and look into clipping thresholds $\mu_{\text{up}} = 0,\: \sqrt{\pi}/10,\: 3\sqrt{\pi}/16$.

We now discuss the variances of the GKP qubits before each GEC step of the repeater protocol. At the beginning of protocol-I, the GKP qubit prepared by Alice has initial quadrature variances $(\sigma^{2}_{q},\: \sigma^{2}_{p} )$.
After both amplification and loss, the variances before the GEC become $(\sigma^{2}_{q} + 2\frac{1- \eta}{\eta},\: \sigma^{2}_{p} + 2\frac{1- \eta}{\eta})$. While performing GEC introduces additional Gaussian noise from the finitely squeezed ancilla states, the logical error rate becomes $E_{X (Z)}(2\sigma^{2} + \frac{1- \eta}{\eta})$ for protocol-I. Similarly, for protocol-II, where the clipping method is applied, the resulting logical error rates are given by $E^{X (Z)}_{\mu_{\text{up}}}(2\sigma^2 + \tfrac{1-\eta}{\eta})$.  
For protocol-III, again while performing imperfect teleportation-based GEC, the logical error rate becomes $E^{X (Z)}_{\mu_{\text{up}}}(2\sigma^{2} +  \frac{(1 - \sqrt{\eta} )}{\sqrt{\eta}} )$.
After GEC round, the variances of the GKP states are reset to $(\sigma^{2}_{q},\: \sigma^{2}_{p} ) = (\sigma^{2},\: \sigma^{2})$.

The results demonstrate a clear advantage in the secure key rate distance as the system progresses from protocol-I to protocol-III (Figures \ref{fig:s1}, \ref{fig:s20}, and \ref{fig:s100}). A primary limitation of standard CV protocols is the rapid degradation of the secure key rate over long distances, often characterised by a sharp ``waterfall". As observed in Figure \ref{fig:s100}, protocol-I experiences this waterfall at approximately 100~km. However, the implementation of protocol-III effectively mitigates this degradation, shifting the waterfall point to twice the distance. 
By integrating clipping into Protocol-II, we’re able to significantly boost the key rates. However, this enhancement is achieved at the cost of a reduced overall success probability. While clipping  amplifies the secure information, it inherently discards a portion of the signals.

Operating with 15 dB of squeezing and a threshold of $\mu_{\text{up}} = \sqrt{\pi}/10$, the system distributes keys over approximately 320 km (Figure \ref{fig:s100}c). Increasing the squeezing level from 15 dB to approximately 18 dB yields an additional 100 km of secure transmission distance. While achieving 18 dB of squeezing pushes the limits of current state-of-the-art quantum optical generation, representing a significant technical hurdle.
Alternatively, the transmission distance can be extended by increasing the number of quantum repeater stations (MQR), while keeping the squeezing constant at 15 dB. Ultimately, the choice between these two scaling methods will depend on the specific constraints of the quantum network, whether it is more feasible to invest in high squeezing technologies or to deploy additional repeater nodes.

\subsection{Optimum of Protocol-III}\label{Optimum of Protocol-III}

\begin{table*}[t]
\centering
\caption{System Optimisation: $r = 15$dB, SKR $\geq 10^{-2}$, $\mu_{\text{up}} = \sqrt{\pi}/10$}
\begin{tabular}{ccccc}
\toprule
\textbf{MQR Dist $L_0$ (km)} & \textbf{Loss/Seg (dB)} & \textbf{Max-MQR} & \textbf{Max-Total-Dist (km)} & \textbf{SKR} \\
\midrule
0.10 km & 0.020 & 32168 & 3216.90 km & $1.000 \times 10^{-2}$ \\
0.20 km & 0.039 & 18413 & 3682.80 km & $1.000 \times 10^{-2}$ \\
0.33 km & 0.065 & 9833  & 3245.22 km & $1.000 \times 10^{-2}$ \\
0.46 km & 0.091 & 5738  & 2639.94 km & $1.000 \times 10^{-2}$ \\
0.59 km & 0.116 & 3592  & 2119.87 km & $1.001 \times 10^{-2}$ \\
0.72 km & 0.142 & 2380  & 1714.32 km & $1.002 \times 10^{-2}$ \\
0.85 km & 0.168 & 1653  & 1405.90 km & $1.001 \times 10^{-2}$ \\
0.98 km & 0.193 & 1193  & 1170.12 km & $1.001 \times 10^{-2}$ \\
1.11 km & 0.219 & 889   & 987.90 km  & $1.002 \times 10^{-2}$ \\
1.24 km & 0.245 & 681   & 845.68 km  & $1.002 \times 10^{-2}$ \\
1.37 km & 0.270 & 534   & 732.95 km  & $1.002 \times 10^{-2}$ \\
1.50 km & 0.296 & 427   & 642.00 km  & $1.003 \times 10^{-2}$ \\
\bottomrule
\end{tabular}
\label{tableI}
\end{table*}

Table~\ref{tableI} presents the optimisation for protocol~III, targeting a secure key rate (SKR) of $\ge 10^{-2}$ at a squeezing parameter of $r =  15\,\mathrm{dB}$. To evaluate the maximum secure range, we systematically vary the physical spacing between repeater stations ($L_0$) and iteratively increase the number of intermediate repeater nodes ($\mathrm{MQR}$) until the accumulated noise degrades the SKR below the target threshold.

The optimisation shows a clear trend: decreasing the distance between neighbouring stations increases the maximum communication distance. For example, when the station spacing is $L_0 = 0.20\,\mathrm{km}$, the loss per segment is only $0.039\,\mathrm{dB}$, which can be effectively corrected by the GKP error-correction scheme. As a result, up to $\mathrm{MQR} = 18{,}413$ repeater nodes can be used, allowing secure communication over a total distance of about $3{,}682\,\mathrm{km}$.

This result suggests that at $r = 15\,\mathrm{dB}$, the 
errors caused by the finite-energy GKP states are already 
well suppressed. Therefore, the best approach is to keep the 
stations close together so that each segment experiences 
very little loss. Note, however that there is a limit to this strategy. If the spacing is decreased to $L_0 = 0.10\,\mathrm{km}$ the maximum distance decreases. At this point the noise introduced by the channel is less than that introduced by the ancilla and further reduction of spacing is counterproductive. 

As the station spacing $L_0$ increases, 
the loss per segment also increases and eventually becomes 
too large to be fully corrected by the recovery operations. 
This reduces the number of repeater nodes that can be 
supported. For example, at $L_0 = 1.50\,\mathrm{km}$, the 
maximum number of nodes drops to $\mathrm{MQR} = 427$, 
limiting the total communication distance to approximately 
$642\,\mathrm{km}$.

The maximum-distance optimisation results are consistent 
with the fixed-node behaviour shown in Fig.~\ref{fig:s100}
(c). The figure illustrates the SKR as a function of 
distance for a network with a fixed maximum quantum repeater 
(MQR) count of $N=100$. For Protocol~III ($\mu_{\mathrm{up}}
=\sqrt{\pi}/10$), the SKR falls below the $10^{-2}$ 
threshold at a total distance of approximately $277\,
\mathrm{km}$. With the number of nodes fixed at $N=100$, the 
inter-node spacing increases to approximately $L_0 \approx 
2.75\,\mathrm{km}$. At this larger spacing, each segment 
experiences greater loss, making it more difficult for the 
finite-squeezing GKP recovery operations to correct the 
accumulated photon-loss errors. As a result, the SKR 
decreases rapidly with distance.
Therefore, the table and figure present complementary 
perspectives. The table shows the maximum achievable 
distances obtained by optimising the repeater spacing, 
whereas the figure illustrates the performance degradation 
that occurs when the number of repeaters is fixed and the 
spacing between them becomes larger.

\section{GKP-Parity-Encoding}\label{GKP-Parity-Encoding}

\begin{figure*}[!htb]
	\centering
		\includegraphics[scale=0.6]{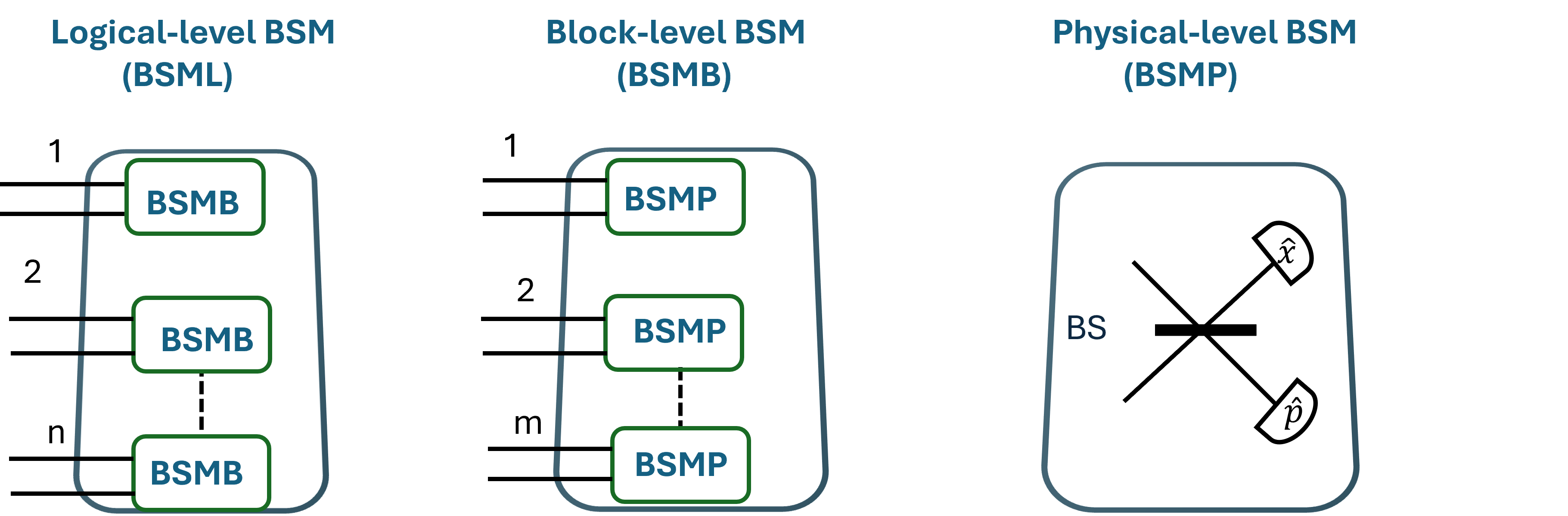}
\caption{ Schematic of CBSM scheme using GKP qubits. The schemes are implemented in a concatenated structure: each logical-level BSM (BSM$\mathrm{L}$) is constructed from a combination of $n$ block-level BSMs (BSM$\mathrm{B}$). In turn, each BSM$\mathrm{B}$ is realized through a combination of $m$ physical-level BSMs (BSM$\mathrm{P}$).
\label{fig:CBSMmain}}
\end{figure*}

\begin{figure*}[!htb]
	\centering
		\includegraphics[scale=0.6]{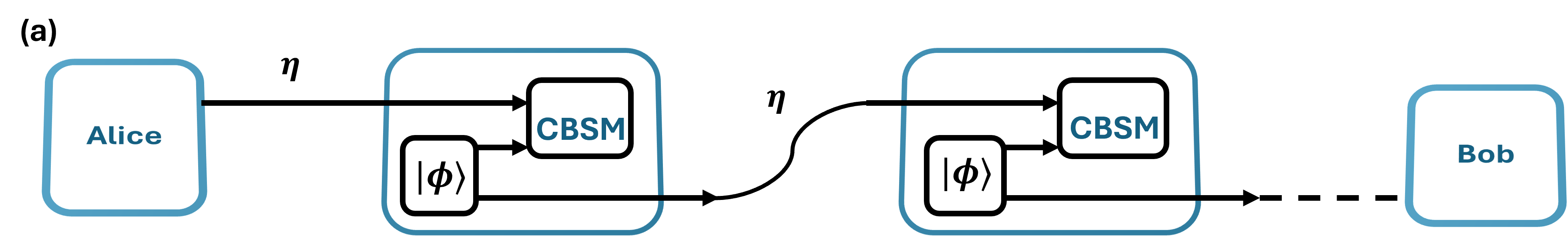}
        \includegraphics[scale=0.6]{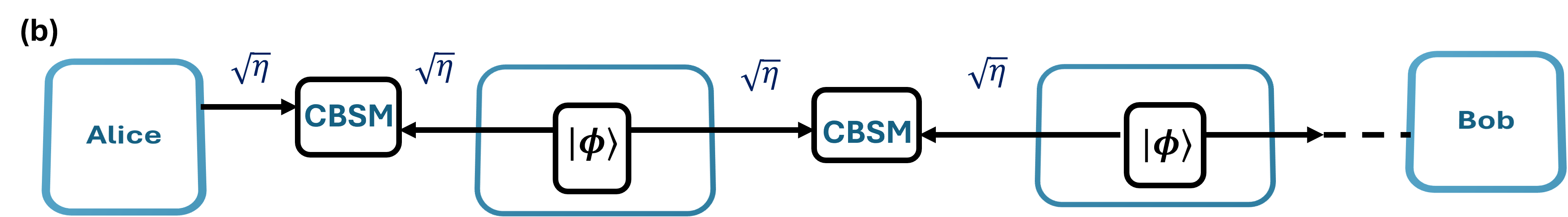}
\caption{ 
Schematic of the quantum repeater architectures. (a) is the standard Bell measurement protocol. %discussed in Section~\ref{BSM: Physical level}. 
The sender (Alice) prepares a parity encoded GKP qubit and transmits it to the first quantum repeater station, where the CBSM protocol is applied. A parity encoded GKP Bell state $\ket{\phi}_+$ is locally prepared, and a CBSM is performed between the incoming qubit and one half of the Bell pair. As a result, the quantum information encoded in the incoming qubit is teleported to the other half of the Bell state, which is subsequently sent to the next repeater station. The CBSM measurement outcomes are transmitted via a classical channel to Bob to enable recovery of the original quantum data. Because of the CBSM protocol, each repeater station can correct logical errors arising from photon loss experienced by the incoming GKP parity qubit.The qubit then propagates through multiple repeater stations separated by a distance $L_{0}$.
(b) is the teleamplifier scheme discussed in Section ~\ref{Teleamplifier in the physical level BSM}. It proceeds in a similar way to the standard scheme except that the CBSM are distributed along the channel at relay points half-way between the repeater stations.
\label{fig:CBSMQR}}
\end{figure*}

As discussed previously, amplification-based strategies for correcting photon loss are generally not optimal. Previously ``tree" encodings have been investigated for quantum communication with GKP states \cite{fukui2021all}. In this section, we focus on Bell-state measurement (BSM) schemes with parity encoding, which have been shown to be optimal for loss correction, particularly in the context of long-distance quantum communication \cite{lee2018fundamental}.
BSM schemes were initially developed for single-photon qubits and coherent-state qubits, employing a single 50:50 beam splitter (BS) and two photon-number detectors \cite{braunstein2005quantum}. These schemes were later extended to concatenated Bell-state measurements (CBSMs) for both single-photon and coherent-state qubits \cite{ralph2005loss, lee2018fundamental, lee2021loss}. In the following, we discuss the CBSM scheme applied to GKP qubits (More details in Appendix.\ref{More details on GKP-Parity-Encoding}). For GKP qubits, the Bell-state measurement can be implemented using a 50:50 beam splitter followed by homodyne detection. In addition, we employ a clipping procedure to keep the GKP states within domain and to discard measurement outcomes associated with unreliable syndrome information. The clipping method is discussed in detail in Section \ref{protocol-II}.

The computational basis states of the ideal GKP qubit are given by
\begin{align}
    \ket{\bar{0}}_{\text{GKP}} & \equiv \sum_{n \in \mathbb{Z}} \ket{\hat{q} = 2n\sqrt{\pi}} \label{zeroideal}\\
    \ket{\bar{1}}_{\text{GKP}} & \equiv \sum_{n \in \mathbb{Z}} \ket{\hat{q} = (2n + 1)\sqrt{\pi}} 
\end{align}

Also, the complementary basis states $\ket{\pm}_{\text{GKP}} \equiv  \frac{1}{\sqrt{2}} (\ket{\bar{0}}_{\text{GKP}} \pm  \ket{\bar{1}}_{\text{GKP}})$ are given by

\begin{align}
   \ket{\bar{+}}_{\text{GKP}} & \equiv \sum_{n \in \mathbb{Z}} \ket{\hat{p} = 2n\sqrt{\pi}} \\
    \ket{\bar{-}}_{\text{GKP}} & \equiv \sum_{n \in \mathbb{Z}} \ket{\hat{p} = (2n + 1)\sqrt{\pi}} 
\end{align}

A GKP Bell pair can be generated directly by applying a balanced beam splitter to two qunaught states \cite{walshe2020continuous} (see Apendix \ref{Background} for more details). We then define the four GKP Bell pairs at each level as

\textbf{Logical level:}
\begin{align}
    \ket{\phi}_{\pm} & \equiv \ket{0}_{\text{L}}\ket{0}_{\text{L}} \pm \ket{1}_{\text{L}}\ket{1} _{\text{L}} \\
    \ket{\psi}_{\pm} & \equiv \ket{0}_{\text{L}}\ket{1}_{\text{L}} \pm \ket{1}_{\text{L}}\ket{0} _{\text{L}}
\end{align}
\begin{align}
    \ket{0}_{\text{L}} & \equiv [\ket{\bar{+}^{\otimes m}}_{\text{GKP}} + \ket{\bar{-}^{\otimes m}}_{\text{GKP}}]^{n} \\
  \ket{1}_{\text{L}}   & \equiv [\ket{\bar{+}^{\otimes m}}_{\text{GKP}} - \ket{\bar{-}^{\otimes m}}_{\text{GKP}}]^{n}
\end{align}

\textbf{Block level:}
\begin{align}
    \ket{\phi^{m}}_{\pm} & \equiv \ket{+^{m}}_{\text{GKP}}\ket{+^{m}}_{\text{GKP}} \pm \ket{-^{m}}_{\text{GKP}} \ket{-^{m}}_{\text{GKP}} \\
    \ket{\psi^{m}}_{\pm} & \equiv \ket{+^{m}}_{\text{GKP}}\ket{-^{m}}_{\text{GKP}} \pm \ket{-^{m}}_{\text{GKP}} \ket{+^{m}}_{\text{GKP}} 
\end{align}

Each logical-level GKP Bell state can be decomposed into $n$ block-level Bell states:
\begin{align}
 \ket{\phi}_{+(-)} & \equiv \frac{1}{\sqrt{2^{n -1}}}\sum_{k = \text{even(odd)} \leq n} \mathcal{P}\Big[\ket{\phi^{m}}_{-}^{\otimes k}\ket{\phi^{m}}_{+}^{\otimes n-k}\Big] \label{LoG1}\\
 \ket{\psi}_{+(-)} & \equiv \frac{1}{\sqrt{2^{n -1}}}\sum_{k = \text{even(odd)} \leq n} \mathcal{P}\Big[\ket{\psi^{m}}_{-}^{\otimes k}\ket{\psi^{m}}_{+}^{\otimes n-k}\Big] \label{LoG2}
\end{align}
Here, $\mathcal{P}[\cdot]$ denotes the sum over all possible permutations of the tensor products enclosed in the square brackets (Detailed calculation in Appendix.\ref{Decomposition of the Encoded Bell States}). Similarly, each block-level Bell state can be decomposed into $m$ physical-level Bell states:
\begin{align}
     \ket{\phi^{m}}_{\pm} & \equiv \frac{1}{\sqrt{2}} \sum_{l = \text{even}\leq m} \mathcal{P}\Big[\ket{\psi}_{\pm}^{\otimes l}\ket{\phi}_{\pm}^{\otimes m-l}\Big] \label{Block1}\\
    \ket{\psi^{m}}_{\pm} & \equiv \frac{1}{\sqrt{2}} \sum_{l = \text{odd}\leq m} \mathcal{P}\Big[\ket{\psi}_{\pm}^{\otimes l}\ket{\phi}_{\pm}^{\otimes m-l}\Big] \label{Block2}  
\end{align}

The outcome is determined by the homodyning results which measures the sum and difference of the quadratures. The four Bell states can be deterministically identified from the homodyning results, except in the case where homodyning fails as clipping fails. 
The beam splitter transforms the quadratures as
\begin{align}
    x'_{1} \to \frac{1}{\sqrt{2}}(x_{1} + x_{2}), \: p'_{1} \to \frac{1}{\sqrt{2}}(p_{1} + p_{2})\\
    x'_{2} \to \frac{1}{\sqrt{2}}(x_{1} - x_{2}), \: p'_{2} \to \frac{1}{\sqrt{2}}(p_{1} - p_{2})
\end{align}
such that homodyning the output modes effectively measures the sum of the x quadratures and the difference of the p quadratures. 

Assuming $(s_p, s_x)$ where $s \in \{\text{even},\text{odd} \}$, we define the following:

\begin{align}
    (\text{even}, 0) & \to \ket{\phi}_{+},\\
     (\text{odd}, 0) & \to \ket{\phi}_{-},\\
      ( 0, \text{even}) & \to \ket{\psi}_{+},\\
      ( 0, \text{odd}) & \to \ket{\psi}_{-}
\end{align}

As realistic GKP states suffer from finite squeezing and transmission noise, these raw homodyne measurements yield continuous, real-valued outcomes rather than perfect logical grid points. To recover the discrete logical information, we apply a clipping scheme, which rounds the raw continuous outcomes to the nearest ideal GKP grid points, effectively filtering out small analog shift errors. We then evaluate the parity of these corrected values to determine if they represent an even or odd multiple of the fundamental grid spacing. The parity of the x-sum distinguishes between the $\ket{\phi}$ and $\ket{\psi}$ Bell states, while the parity of the p-difference determines the relative phase ($+$ or $-$). Consequently, the four Bell states are uniquely and deterministically identified by these discrete parity pairs. If the accumulated analog shift error pushes the raw measurement too close to the midpoint between two valid grid bins, the clipping procedure becomes ambiguous, and the Bell-state measurement is flagged as a failure.

Each logical-level BSM is implemented as a composition of $n$ block-level BSMs, and each block-level BSM is, in turn, realized as a composition of $m$ physical-level BSMs.
The core structure of the CBSM is summarized in Eqs. (\ref{LoG1}), (\ref{LoG2}), (\ref{Block1}), and (\ref{Block2}). These equations enable the realization of a logical BSM through the combination of $n$ block-level BSMs, each of which is further constructed from $m$ physical-level BSMs. They show that, at the outset, a CBSM does not introduce any logical errors; the only possible outcomes are success or failure. The present work considers approximate GKP states, when the scheme succeeds (or when loss is detected and or a shift error occurs), both the photon loss and finite squeezing contributes to the logical errors.
We now describe the different layers of the scheme to justify the construction presented in Eqs. (\ref{LoG1}), (\ref{LoG2}), (\ref{Block1}), and (\ref{Block2}).

\subsection{BSM: Physical level }
\label{BSM: Physical level}

For the Bell-state measurement (BSM) on GKP Bell pairs, we employ the standard scheme consisting of beam splitter (BS) and homodyne measurement. This setup implements a CV Bell measurement allows for the deterministic identification of the four Bell states, namely $ \ket{\phi}_{\pm}$ and $ \ket{\psi}_{\pm}$. We interface the homodyne readout via a clipping protocol. By mapping results to their nearest lattice points, this post-selection process ensures that only the most reliable qubits are kept. If the homodyne readout falls significantly outside the central regions of the lattice, the event is discarded as a BSM failure. When the clipping results are ambiguous, the measurement may fail to discriminate full Bell states. Specifically, where only the relative sign of the Bell state (i.e., the “$\pm$” in $ \ket{\psi}_{\pm}$ or $ \ket{\phi}_{\pm}$) remains resolvable, while the which-pair parity ($ \ket{\phi}_{+}$ and $ \ket{\psi}_{+}$) is lost.

\subsection{BSM: Block level}
\label{BSM: Block level}

A block-level BSM is performed by carrying out physical-level BSMs on each mode within the block. The sign of the block-level Bell state is determined by a majority vote over the signs obtained
from the physical-level BSM outcomes. The Bell-state letter is determined by the parity of the 
number of physical-level BSM results labelled with the $\psi$ symbol: the block-level state is $
\phi$ ($\psi$) if this number is even (odd). In contrast, the Bell-state letter cannot be 
determined if at least one physical-level BSM fails, in which case we regard the block-level BSM
as having failed. 

\subsection{BSM: Logical level }
\label{BSM: Logical level}
A logical-level BSM is performed by executing block-level BSMs on each block. The sign of the logical-level Bell state is determined by the parity of the number of block-level BSM outcomes with a minus sign: it is plus (minus) if this number is even (odd). The Bell-state letter is determined by a majority vote over the letters of the block-level BSM results, excluding any that have failed. As before, the sign of the logical-level Bell state is always well defined. However, the letter cannot be determined if all block-level BSMs fail, or if the resulting block-level Bell states have an equal number of both letters. In such cases, we regard the logical-level BSM as having failed. In the ideal, error-free scenario, these failure modes vanish entirely; all block-level BSMs succeed and unanimously agree on the Bell-state letter, rendering the logical measurement perfectly deterministic.\\

\subsection{The teleamplifier CBSM scheme}
\label{Teleamplifier in the physical level BSM}

In the standard approach the CBSM occur in the repeater station where the Bell resource states are produced. However, we can take advantage of the teleportation feature of the error correction to distribute the CBSM between the repeater stations in a similar way as protocol III in the unencoded case. We refer to this as the teleamplifier scheme.
%For Bell-state measurements (BSMs) on GKP Bell pairs, we adopt an integrated Bell measurement scheme that directly incorporates relay-style amplification (teleamplification), rather than relying on the conventional beam splitter and homodyne detection approach. This method introduces an additional round of teleportation-based error correction at each repeater station, achieved by embedding an extra amplifier within the resource state. 
As a result, both modes undergo equal loss but are transmitted over only half the link distance, $L_0/2$. Consequently, each loss channel is characterized by a transmission efficiency of $\sqrt{\eta}$. By embedding teleamplification directly into the Bell measurement stage, this method compensates for channel loss during transmission while preserving the deterministic discrimination of the four Bell states, namely $\ket{\phi}_{\pm}$ and $\ket{\psi}_{\pm}$. The homodyne readout is interfaced through a clipping protocol, where measurement outcomes are mapped to their nearest lattice points to retain only the most reliable qubits. Events with outcomes significantly outside the lattice regions are discarded as BSM failures, while ambiguous clipping results may still lead to partial Bell-state discrimination, where only the relative sign (“$\pm$”) remains identifiable and the full parity information is lost. In the ideal limit of infinite squeezing and noiseless transmission, such failures vanish entirely, and the integrated amplification enhanced GKP BSM yields perfectly consistent outcomes. With this modification at the physical level BSM, the block-level and logical-level BSMs remain unchanged, as described in subsections~\ref{BSM: Block level} and \ref{BSM: Logical level}.

As discussed in Appendix~\ref{Simulation details of the GKP-parity-encoding}, we follow the same simulation strategy, incorporating transmission losses $\eta_{0}$ and $\eta_{L_{0}}$, with the following total effective Gaussian noise variance:
\begin{equation}
\sigma_{\mathrm{tot}}^2(\eta_{0},\eta_{L_{0}}) =
2\sigma_{\mathrm{GKP}}^2
+ \frac{1-\sqrt{\eta_{0}}}{\sqrt{\eta_{0}}}
+ \frac{1-\sqrt{\eta_{L_{0}}}}{\sqrt{\eta_{L_{0}}}}.
\end{equation}

\subsection{One-way repeater protocol}

The overall one-way repeater protocols for transmitting quantum information along a network, are illustrated in Fig. \ref{fig:CBSMQR}. In each building block of this scheme, a CBSM is performed on two GKP qubits. For the standard case one qubit propagates between repeater nodes (over a distance $L_{0}$), while the other is detected within the repeater. The success probability of this operation is given by $P(\eta_{L_{0}}, \eta_{0})$, where $\eta_{0}$ denotes the effective loss rate inside the repeater, and $\eta_{L_{0}}$ represents the effective loss rate of the travelling qubit over distance $L_0$. These two parameters are related as  $\eta_{L_{0}} = \eta_{0} e^{-L_0/ L_{\text{att}}}$ with the attenuation length $L_{\text{att}} = 22$ km.
In this one-way communication scheme, the total distance $L$ between Alice and Bob is divided into segments of length $L_0$, which determines the number of repeater nodes required. Our repeater model consists of two main components: one for preparing a logical Bell pair, and another for performing CBSM. Importantly, it does not require long-lived quantum memories, photon-matter interactions, or complex circuit operations. %At each repeater node, CBSM is applied between the incoming GKP qubit and one qubit from the resource GKP pair.
In the teleamplifier scheme the propagation distance to the CBSM is effectively halved, as discussed in the previous section. The effective transmission is only one criterion for a successful building block; we also account for the post-selection clipping method. Since GKP states are prone to logical errors when displacement noise pushes the state near the boundaries of the lattice cells, measurement outcomes falling into these ambiguous regions are discarded. By imposing a clipping threshold, accepting only outcomes within a trusted window around the ideal lattice points, we discard unreliable syndrome information. Consequently, the success probability $P(\eta_{L_{0}}, \eta_{0})$ represents that the qubit survives the effective loss channel and satisfies this clipping threshold.

To evaluate the performance of the GKP-parity-encoded quantum repeater, we compute the asymptotic secure key rate (SKR), which is governed by both the residual logical error rate and the end-to-end success probability of the protocol. Assuming a symmetric noise channel where logical bit-flip and phase-flip errors are equivalent ($E_{X(Z)}$), the secret key rate per attempted logical transmission is defined as $\mathrm{SKR} = P_{\text{suc}} \times \max{(0, 2h(E_{X(Z)}))}$, where $h(\cdot)$ represents the binary Shannon entropy and $P_{\mathrm{suc}} = (P(\eta_{L_{0}}, \eta_{0}))^{L/L_0}$ is the probability not to fail during the entire transmission. Our results in Subsection~\ref{Calculation of the cost} demonstrate that, by optimising $m$, $n$, and the spacing between repeater stations ($L_0$), we can identify optimal architectures that minimise logical error probability while maintaining high success probability, thereby maximising the achievable secure communication range.

\subsection{Calculation of the cost}
\label{Calculation of the cost}
We consider the resource cost (RC) \cite{munro2012quantum}, defined as the average total number of qubits consumed in the repeater protocol. To optimize the transmission of a logical qubit over a distance $L$, we perform numerical searches to identify the optimal parameters. The optimized parameters, denoted as $RC(n, m, L_{0})$, are chosen to minimize the total qubit cost $RC$, taking into account possible losses affecting both qubits during transmission and operations within the repeater. The minimized resource cost $RC$ is

\begin{align}
    RC_{\text{min}} & = \min_{n, m,  L_{0}} RC(r, n, m, \mu_{\text{up}},  L_{0})
\end{align}

Each logical-level BSM is implemented through a composition of $n$ block-level BSMs, and each block-level BSM, in turn, is composed of $m$ physical-level BSMs. The parameter $L_{0}$ corresponds to the distance between repeater nodes. 

We consider several transmission distance scenarios while keeping the squeezing fixed at $r = 14.0\,\mathrm{dB}$ and $\mu_{\text{up}} = 0.15$. The results for the standard Bell measurement at the physical level CBSM (\ref{BSM: Physical level}) and the CBSM with teleamplifier at the physical level (\ref{Teleamplifier in the physical level BSM}) are presented as follows.\\

(i) for 1000 KM,\\
\textbf{Standard CBSM:}
$\eta_{0} = 0.98$, $RC_{\text{min}}  = 1.750 \times 10^5$, SKR = 0.8901,  m = 7, n = 15, $L_{0} = 1.20$ KM.\\

\textbf{Teleamplifier CBSM:}
$\eta_{0} = 0.98$, $RC_{\text{min}}  = 7 \times 10^4$, SKR = 0.9833,  m = 7, n = 15, $L_{0} = 3.0$ KM.\\

(ii) for 5,000 KM,\\
\textbf{Standard CBSM:}
$\eta_{0} = 0.98$, $RC_{\text{min}}  = 8.750 \times 10^5$, SKR = 0.6052,  m = 7, n = 15, $L_{0} = 1.20$ KM.\\

\textbf{Teleamplifier CBSM:}
$\eta_{0} = 0.98$, $RC_{\text{min}}  = 3.5 \times 10^5$, SKR = 0.9329,  m = d7, n = 15, $L_{0} = 3.0$ KM.\\

(iii) for 10,000 KM,\\
\textbf{Standard CBSM:}
$\eta_{0} = 0.98$, $RC_{\text{min}}  = 1.750 \times 10^6$, SKR = 0.3504,  m = 7, n = 15, $L_{0} = 1.20$ KM. \\

\textbf{Teleamplifier CBSM:}
$\eta_{0} = 0.98$, $RC_{\text{min}}  = 6.067 \times 10^5$, SKR = 0.6071,  m = 7, n = 13, $L_{0} = 1.20$ KM.\\

A comparison at a transmission distance of $5{,}000\,\mathrm{km}$ demonstrates the higher efficiency of the teleamplifier-based CBSM scheme. Compared with the standard CBSM, it lowers the minimum resource cost ($\mathrm{RC}_{\min}$) by approximately $525{,}000$ qubits. In addition, this improvement is achieved while allowing a significantly larger repeater spacing ($L_0 = 3.0\,\mathrm{km}$ compared to $1.20\,\mathrm{km}$) and maintaining 54\% higher SKR.

In the current experimental setup, selecting m = 3 and n = 3 highlights the limitations of low block sizes in the GKP-parity-encoding scheme. Specifically, with small n and m values lead to inefficient logical encoding, where the error-correction capability becomes insufficient to maintain high transmission rate. For instance, even for a relatively short transmission distance of 100 km, implementing GKP-parity-encoding with teleamplifier would require parameters such as $\eta_{0} = 0.98$, $RC_{\text{min}}  = 1500$, $L_{0} = 1.20$ KM with squeezing $r = 14.0$dB and obtain SKR = 0.7974. The substantial resource cost further emphasises that the protocol illustrated in Fig.~\ref{fig:Stg123} offers a more practical approach for achieving higher transmission rates across shorter distances.

To validate our design, we compare it with two architectures: Lee's photon-qubit parity encoding \cite{lee2018fundamental} and a GKP-based BSM using PNRD (discussed in Appendix \ref{More details on GKP-Parity-Encoding}). We observe that CBSM with CV Bell measurement requires significantly fewer qubits, since GKP qubits are intrinsically protected against errors from the start, unlike single-photon qubits. Specifically, compared with Lee’s photonic version of parity encoding, transmitting over 5000 km using the GKP parity encoding with teleamplifier requires approximately 650,000 fewer qubits. In contrast, Lee’s scheme requires about $1.0 \times 10^{6}$.
We anticipate that this extended code will be effective in practical scenarios, as it is capable of correcting logical errors arising from both photon loss and finite squeezing.

\section{Discussion and Conclusion}
We have proposed a one-way quantum repeater protocol based on GKP qubits to enable long-distance quantum communication. The protocol advances through three progressive protocols, each designed to correct photon loss and shift errors effectively. We began by establishing a base layer where standard teleportation-based GKP error correction and amplification are performed at every repeater station to ensure basic signal recoverability. Building upon this, we integrated a clipping method that employs post-selection to discard unreliable qubits based on measurement outcomes, significantly enhancing the quality of the remaining ensemble. Finally, the architecture is augmented with a relay-like amplification, where resource states are prepared with an effective transmission efficiency of $\sqrt{\eta}$ to realize virtual in-line amplification. By systematically combining teleportation-based correction, clipping, and relay-like amplification, our proposed approach offers a viable pathway for realizing  high communication rates for long distances using GKP qubits. Furthermore, all these protocols can be implemented at room temperature. Also, simple entangled states such as Bell pairs can be generated from product states using passive beam-splitter transformations. 

However, as we established in Subsection~\ref{Optimum of Protocol-III} (see Table~\ref{tableI}) that Protocol~III achieves its best performance when repeater stations are placed as closely together as possible, minimising per-segment loss so that GKP error correction can operate near its effective limit. Therefore, to push the communication range to thousands of kilometres while preserving a higher secure key rate, we transition to investigating the use of GKP-parity encoding.

We have further investigated the use of GKP encoded qubits as a resource for long-distance quantum communication, focusing on a concatenated Bell-state measurement (CBSM) scheme utilizing a 50:50 beam splitter followed by homodyne detection. A central feature of our approach is the integration of a clipping method by discarding measurement outcomes associated with unreliable syndrome. We also investigate the CBSM scheme with a teleamplifier at the physical level.
Our results show that, relative to Lee’s photonic parity-encoding approach \cite{lee2018fundamental}, the GKP parity-encoding scheme with a teleamplifier dramatically reduces resource requirements for transmission over 5000 km, using approximately 65\% fewer qubits. By comparison, Lee’s scheme requires roughly ($1.0 \times 10^{6}$) qubits. Furthermore, we evaluated our homodyne-based approach against a GKP-based scheme utilising Photon Number Resolving Detectors (PNRD) (detailed in Appendix \ref{Alternative way of GKP-parity-encoding}). Our analysis confirms that the proposed CBSM with teleamplifier significantly outperforms these alternatives. It requires fewer resource qubits and offers a practical pathway for long-distance quantum communication. 

The proposed parity-encoded GKP repeater significantly enhances secret key generation while maintaining resource requirements comparable to the existing GKP-based repeater protocols. For example, the scheme of Fukui \emph{et al.} \cite{fukui2021all}, which is based on Varnava's code \cite{varnava2006loss}, requires an average total of $2.2 \times 10^{5}$ GKP qubits to achieve a communication distance of $5000\,\mathrm{km}$ with a key rate on the order of $10^{-3}$ at a squeezing level of $15\,\mathrm{dB}$. By comparison, our parity-encoding scheme achieves the same communication distance using only $1.833 \times 10^{5}$ GKP qubits, representing a reduction of approximately $16.7\%$ in the required number of GKP qubits. At the same time, it achieves a key rate on the order of $10^{-1}$, corresponding to a two-orders-of-magnitude ($100\times$) improvement over the key rate reported in Ref.~\cite{fukui2021all}.

In parity encoding with photonic polarization qubits, the use of PNRDs at each level makes the process inherently nondeterministic. In contrast, GKP-based parity encoding relies on homodyne measurements, which are deterministic in nature. This removes the probabilistic bottleneck during the encoding steps themselves. However, this advantage comes with the challenge of preparing GKP states, which remains experimentally demanding.

By effectively converting the analog information from homodyne detection into reliable digital syndromes, our method bridges the gap between ideal theoretical models and realistic experimental conditions. This establishes the homodyne-based CBSM as a resource-efficient candidate for future optical quantum information processing.

For a transmission distance of approximately $\sim$4000\,km, the choice ultimately rests 
with the experimenter -- one may opt for a dense physical infrastructure of $\sim$18,000 relay nodes using GKP error correction to achieve a key rate on the order of $10^{-2}$, or alternatively, embrace a more qubit-intensive but architecturally compact GKP parity encoding scheme with just $m = 5$ physical repeater nodes and $n = 9$ block-level encoding, investing roughly $1.2 \times 10^{5}$ qubits in exchange for a higher secret key rate of $0.8$ with distance between nodes of 3km-- a compelling reminder that in quantum repeater design, the frontier lies not in a single optimal solution, but in navigating the rich trade space between physical network density and logical encoding overhead.

The most challenging aspects of realizing our scheme lie at the physical level, the generation of GKP states required for physical-level Bell-state measurements. Fortunately, a variety of suitable implementation methods have been proposed, even though some remain nondeterministic or resource-intensive. Inevitably, experimental imperfections in these physical-level processes will affect the overall performance of the scheme, and a detailed investigation of such effects is left for future work.

\section{Acknowledgement}
SNS appreciates Anand Dev Ranjan’s help in creating the circuits.
SNS was supported by the Sydney Quantum Academy, Sydney, NSW, Australia. This work was
partially supported by the Australian Research Council Centre of Excellence for Quantum Computation and
Communication Technology (Project No.CE110001027).

\begin{widetext}
\appendix
\section{Background}
\label{Background}
\subsection{Gottesman-Kitaev-Preskill qubit}
\label{Gottesman-Kitaev-Preskill qubit}
\subsubsection{Ideal GKP qubit}
Let $\hat{q} = (\hat{a} + \hat{a}^{\dag})$ and $\hat{p} = i(  \hat{a}^{\dag} - \hat{a})$ be the position and momentum operators of a bosonic mode, where $\hat{a}$ and $\hat{a}^{\dag}$ are annihilation and creation operators satisfying $[\hat{a}, \hat{a}^{\dag}] = 1$. We define the GKP qubit as the two-dimensional subspace of a bosonic Hilbert space that is stabilized by the two stabilizers
\begin{align}
    \hat{S}_{q} & = Z(2\sqrt{\pi}) \equiv \exp(i 2\sqrt{\pi}\hat{q}) \\
     \hat{S}_{p} & = X(2\sqrt{\pi}) \equiv \exp( - i 2\sqrt{\pi}\hat{p})
\end{align}
In general, phase-space displacement operators are defined as
\begin{align}
    X(s) & = \exp( - i s\hat{p}) \\
    Z(s) & = \exp(  i s\hat{q})
\end{align}
where their actions on position eigenstates are given by
\begin{align}
    X(s)\ket{q} & = \ket{q + s} \\
    Z(s)\ket{q} & = e^{i sq}\ket{q}
\end{align}
where $X(s)$ is a displacement in position by $q$. $Z(s)$ is a displacement in momentum by p.
Measuring these two commuting stabilizers is equivalent to measuring the position and momentum operators $\hat{q}$  and $\hat{p}$ modulo $\sqrt{\pi}$.
Therefore, any phase-space shift error $\exp{i(\varsigma_{p} \hat{q} - \varsigma_{q} \hat{p})}$ acting on the ideal GKP qubit can be detected and corrected as long as $|\varsigma_{q}|$, $|\varsigma_{p}|$ $< \frac{\sqrt{\pi}}{2}$.
Explicitly, the computational basis states of the ideal GKP qubit \cite{PhysRevA.64.012310} are given by
\begin{align}
    \ket{\bar{0}}_{\text{GKP}} & \equiv \sum_{n \in \mathbb{Z}} \ket{\hat{q} = 2n\sqrt{\pi}} \\
    \ket{\bar{1}}_{\text{GKP}} & \equiv \sum_{n \in \mathbb{Z}} \ket{\hat{q} = (2n + 1)\sqrt{\pi}} 
\end{align}

Also, the complementary basis states $\ket{\pm}_{\text{GKP}} \equiv  \frac{1}{\sqrt{2}} (\ket{\bar{0}}_{\text{GKP}} \pm  \ket{\bar{1}}_{\text{GKP}})$ are given by

\begin{align}
   \ket{\bar{+}}_{\text{GKP}} & \equiv \sum_{n \in \mathbb{Z}} \ket{\hat{p} = 2n\sqrt{\pi}} \\
    \ket{\bar{-}}_{\text{GKP}} & \equiv \sum_{n \in \mathbb{Z}} \ket{\hat{p} = (2n + 1)\sqrt{\pi}} 
\end{align}
Clearly, all these basis states have $\hat{q} = \hat{p} = 0$ modulo $\sqrt{\pi}$ and thus are stabilized by $\hat{S}_{q}$ and $ \hat{S}_{p}$.

\subsubsection{Approximate GKP states}
In an experimental setting one will be dealing with finitely squeezed states. The canonical way to generate approximate GKP states is to replace the delta functions with Gaussians of width $\Delta$ and then introduce an overall Gaussian envelope of width $\kappa^{-1}$:
\begin{align}
   \ket{0}_{\Delta, \kappa} \propto \sum_{n \in \mathbb{Z}} e^{-\frac{1}{2}\kappa^{2}(2n\sqrt{\pi})^{2}} X(2n\sqrt{\pi}) \ket{\Delta}_{q} \label{zerogkpphy}
\end{align}

where $\ket{\Delta}_{q}$ is defined as
\begin{align}
   \ket{\Delta}_{q} & = \int \,dq (\frac{1}{\pi \Delta^{2}})^{1/4} e^{\frac{-q^{2}}{2 \Delta^{2}}} \ket{q}
\end{align}
$\ket{0}_{\Delta, \kappa}$ becomes,
\begin{align}
   \ket{0}_{\Delta, \kappa} \equiv (\frac{1}{\pi \Delta^{2}})^{1/4} \sum_{n \in \mathbb{Z}} e^{-\frac{1}{2}\kappa^{2}(2n\sqrt{\pi})^{2}} e^{\frac{-q^{2}}{2 \Delta^{2}}} \ket{q + 2n\sqrt{\pi} } \label{ApproxzeroGKP}
\end{align}

Similarly,
\begin{align}
   \ket{1}_{\Delta, \kappa} & \propto \sum_{n \in \mathbb{Z}} e^{-\frac{1}{2}\kappa^{2}((2n + 1)\sqrt{\pi})^{2}} X((2n + 1)\sqrt{\pi}) \ket{\Delta}_{q} \\
   & \propto (\frac{1}{\pi \Delta^{2}})^{1/4} \sum_{n \in \mathbb{Z}} e^{-\frac{1}{2}\kappa^{2}((2n + 1)\sqrt{\pi})^{2}} e^{\frac{-q^{2}}{2 \Delta^{2}}} \ket{q + (2n + 1)\sqrt{\pi} }
\end{align}

This is just one method to transition from infinite to finite energy states \cite{tzitrin2020progress}. However, three conventional approximations of the GKP codes were analytically shown to be equivalent in \cite{matsuura2020equivalence}.
It is common to work in the regime where $\Delta = \kappa$, so that
in the $\Delta \to 0$ limit the number of spikes increases while the
width of each spike decreases, and the normalizable states
approach the ideal ones. In this case,
\begin{align}
    \ket{\psi}_{\Delta} = \ket{\psi}_{\Delta, \kappa=\Delta}
\end{align}

A less cumbersome way than Eq.\ref{zerogkpphy} of expressing $\ket{\psi}$, as pointed out in \cite{menicucci2014fault, noh2018quantum}, is to apply a nonunitary envelope  operator $E(\epsilon) = e^{-\epsilon \hat{n}}$ to the ideal state, where $\hat{n} \equiv \hat{a}^{\dag}\hat{a} = (\hat{q}^{2} + \hat{p}^{2})$ is the number operator:

\begin{align}
    \ket{\psi}_{\epsilon} = E(\epsilon) \ket{\bar{\psi}} \label{epsilongkp}
\end{align}

One can think of this operator as being the result of interfering an ideal GKP state and a vacuum state at a beam splitter, measuring one of the modes, and postselecting on the vacuum \cite{noh2020encoding}.
Now, applying a Gaussian operation to a normalizable state can be viewed as applying this operation to an ideal state followed by a modified envelope:

\begin{align}
    U\ket{\psi}_{\epsilon} & = UE(\epsilon) \ket{\bar{\psi}} \\
    & = \tilde{E}(\epsilon)U\ket{\bar{\psi}} \\
   \tilde{E}(\epsilon) & = U  E(\epsilon) U^{\dag}
\end{align}

\subsection{Teleportation-based GKP Error Correction}
\label{Teleportation-based GKP Error Correction}

\subsubsection{Additional calculation on Teleportation-based GKP error correction using Ideal GKP states}
As the teleportation-based GKP error correction is based on beam-splitter interactions. Under the beamsplitter interaction $\hat{B}_{k \to l}(\pi/4)$, position eigenstates are transformed as
 \begin{align}
     \hat{B}_{k \to l}\ket{ \hat{q_k} = q_k}\ket{\hat{q_l} =q_l} = \ket{\frac{q_k - q_l}{\sqrt{2}}}  \ket{\frac{q_k + q_l}{\sqrt{2}}}
 \end{align}

The so-called qunaught state is defined as

\begin{align}
    \ket{\bar{\O}}_{\text{GKP}} & \equiv \sum_{n \in \mathbb{Z}} \ket{\hat{q} = n\sqrt{2\pi}} \\
    & \equiv \sum_{n \in \mathbb{Z}} \ket{\hat{p} = n\sqrt{2\pi}}
\end{align}

which is useful for quantum sensing applications \cite{duivenvoorden2017single}. This state exhibits a $\sqrt{2\pi}$ periodicity in both position and momentum. Applying the balanced beam-splitter interaction to two GKP qunaught states, we get a GKP-Bell state.

\begin{align}
   \ket{\psi'}  \equiv & \hat{B}_{23}(\pi/4)\ket{\O }_{2}\ket{\O }_{3}\\
   = &  \hat{B}_{23}\sum_{n_2, n_3 \in \mathbb{Z}} \\ 
= &   \sum_{n_2, n_3 \in \mathbb{Z}} \ket{\hat{q}_2 \equiv (n_2 - n_3)\sqrt{\pi}} \ket{\hat{q}_3 \equiv (n_2 + n_3)\sqrt{\pi}}  \\
 = & \ket{0}_2\ket{0}_3 + \ket{1}_2\ket{1}_3 \equiv \ket{\phi^{+}}_{23}
\end{align}

Then, applying a balanced beam-splitter interaction $\hat{B}_{1 \to 2}(\pi/4)$, we find

\begin{align}
    \ket{\psi''} & \equiv  \hat{B}_{12}(\pi/4)  \ket{\psi}_{1}\ket{\phi^{+}}_{23}   \\
   & =     \hat{B}_{12}  \ket{\psi}_{1}( \ket{0}_2\ket{0}_3 + \ket{1}_2\ket{1}_3 )\\
   & = \sum_{n \in \mathbb{Z}} \int dq_1 \psi(q_1) \ket{\hat{q}_1 =\frac{ q_1 - q_2}{\sqrt{2}}}   \ket{\hat{q}_2 =\frac{ q_1 + q_2}{\sqrt{2}}} \ket{\hat{q}_3} \\
   & = \sum_{n \in \mathbb{Z}} \int dq_1 \psi(q_1) \ket{\hat{q}_1 =\frac{ q_1 - n\sqrt{\pi}}{\sqrt{2}}}   \ket{\hat{q}_2 =\frac{ q_1 + n\sqrt{\pi}}{\sqrt{2}}} \ket{\hat{q}_3} \\
   & = \sum_{n \in \mathbb{Z}} \int dq_1 \psi(q_1) \ket{\frac{ q_1 }{\sqrt{2}}-\frac{ n\sqrt{\pi}}{\sqrt{2}}}   \ket{\frac{ q_1 }{\sqrt{2}} + \frac{ n\sqrt{\pi}}{\sqrt{2}}} \ket{\hat{q}_3} 
\end{align}

Thus, after measuring $\hat{q}_1 = q_m$ and $\hat{p}_2 = p_m$ via homodyne measurements, we get

\begin{align}
    \ket{\psi (q_m, p_m)} & \equiv \langle q_m | p_m | \psi'' \rangle \\
    & = \psi(\sqrt{2}q_m + n \sqrt{\pi}) \exp(-ip_m (q_m + n\sqrt{2 \pi})) \ket{q_3}\\
    & = \exp(-ip_m (q_m + n\sqrt{2 \pi})) \ket{q_3} \langle \sqrt{2}q_m + n \sqrt{\pi} | \psi \rangle \\
    & = \ket{q_3 \equiv n\sqrt{\pi}} \bra{q_3 \equiv n\sqrt{\pi}} \exp(-ip_m (q_m + n\sqrt{ \pi})) \nonumber \\
    & \exp( i \sqrt{2} q_m ) \ket{\psi} \\
    & = \ket{q_3 \equiv n\sqrt{\pi}} \bra{q_3 \equiv n\sqrt{\pi}} \exp(-ip_m q_m ) \nonumber\\
    &\exp(-ip_m n\sqrt{2 \pi} ) \exp(-i\sqrt{2}q_m ) \ket{\psi} \\
    & = \hat{\pi}_m \hat{D}(-\sqrt{2}(q_m + p_m))\ket{\psi}
\end{align}

\begin{figure}[!htb]
	\centering
		 \includegraphics[width=0.5\linewidth]{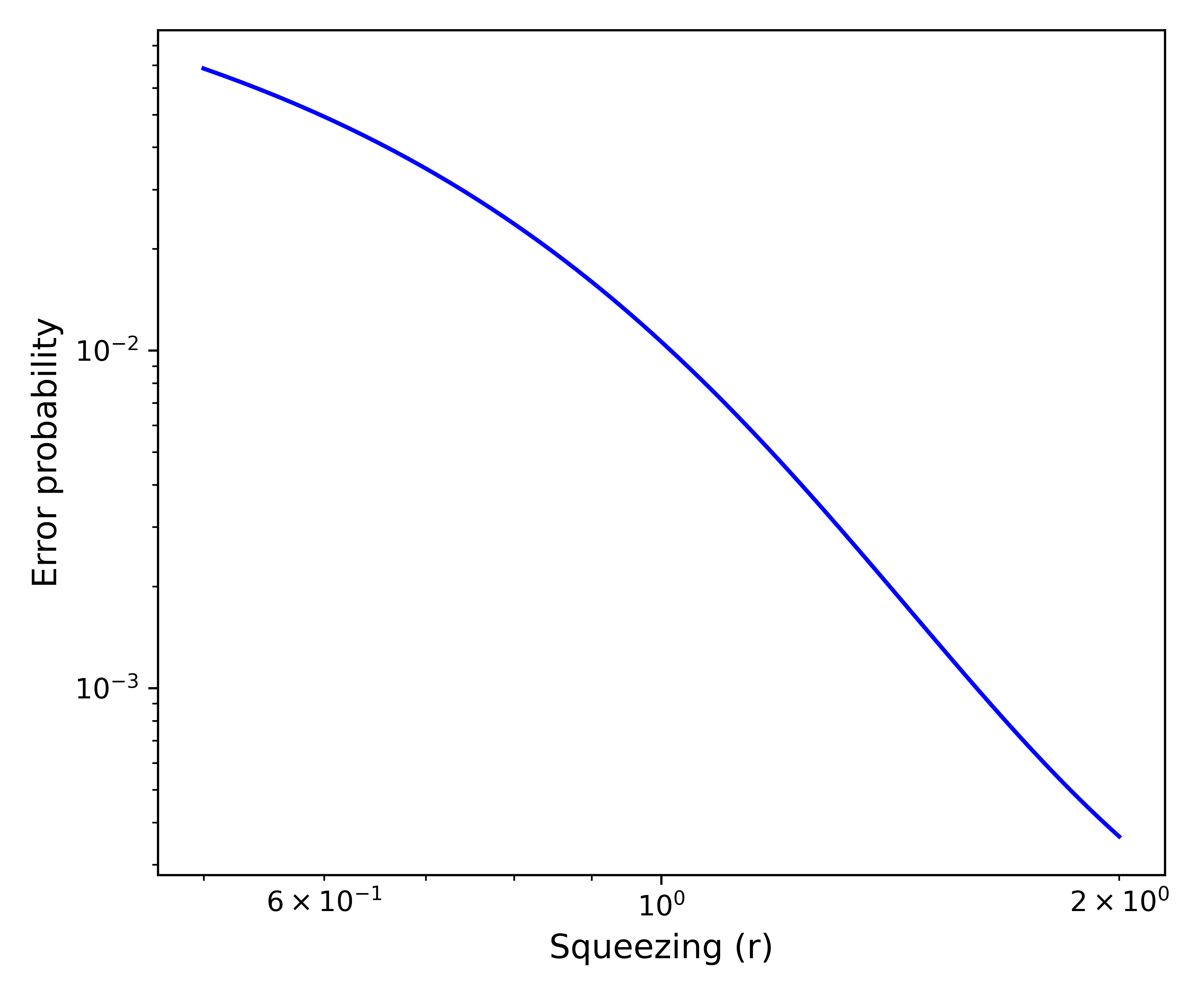}
		\caption{ 
        The error probability of the GKP qubit as a function of the squeezing factor $r$, while undergoing perfect teleportation-based GEC and amplification. Here we consider $\eta = 0.95$.
  \label{ERPROB1}}
\end{figure}

\begin{figure*}[!htb]
	\centering
		\begin{subfigure}{\columnwidth}
             \caption{ \label{plot:error}}
			\includegraphics[width=0.5\columnwidth]{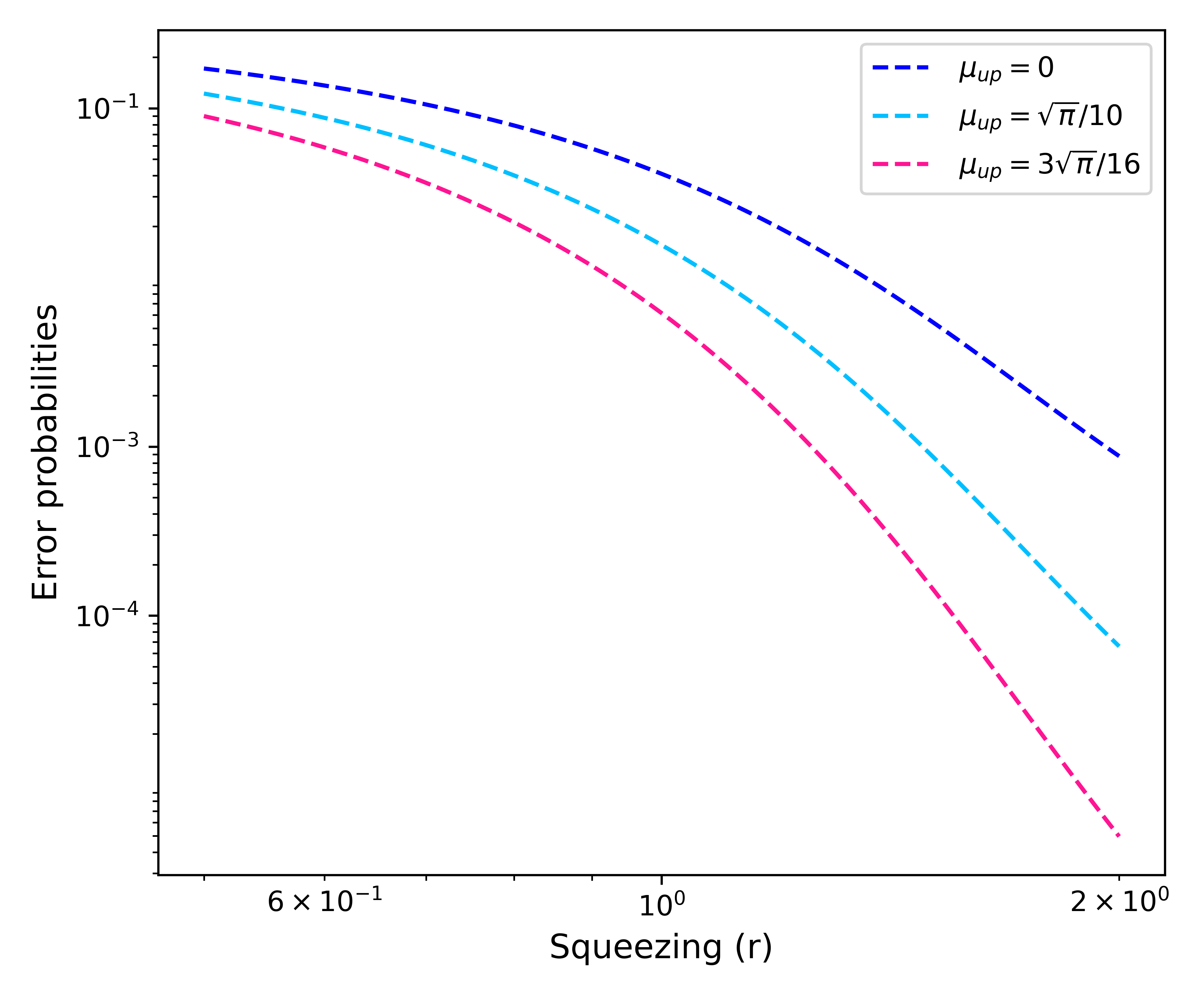}
			
		\end{subfigure}
  		\begin{subfigure}{\columnwidth}
              \caption{ \label{plot:succ}}
			\includegraphics[width=0.5\columnwidth]{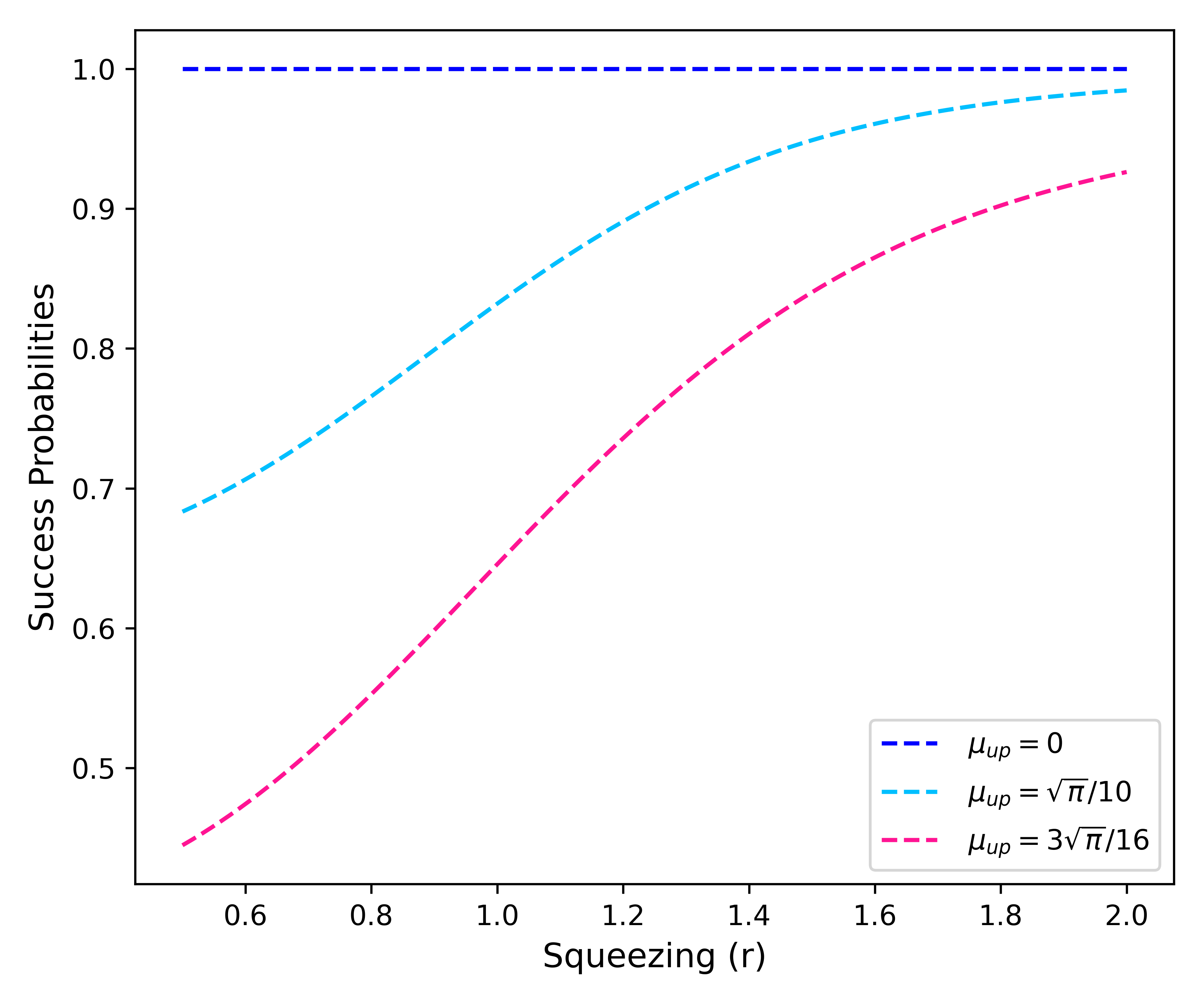}
		\end{subfigure}
		\caption{ (a) The probability of incorrectly identifying the bit value using the clipping method.
(b) The success probability of the clipping method. The dashed blue line ($\mu_{\text{up}} = 0$) corresponds to the error probability of the GKP qubit while performing imperfect teleportation-based GEC and amplification without applying the clipping method, so that the success probability remains unity. The dashed deep sky-blue and deep pink lines depict the error probabilities with imperfect GEC and amplification combined with clipping at $\mu_{\text{up}} = \sqrt{\pi}/10$ and $\mu_{\text{up}} = 3\sqrt{\pi}/16$, respectively. These dashed deep sky-blue and deep pink lines also indicate the success probabilities of the clipping method in (b).
  \label{fig:Clipping}}
\end{figure*}

To correct shift errors in the GKP code, it is crucial to nondestructively measure quadrature operators modulo an appropriate spacing (for example, $\hat{q}$ modulo $\sqrt{\pi}$ and $\hat{p}$ modulo $\sqrt{\pi}$ in the case of the square-lattice GKP code). In the original GKP proposal \cite{PhysRevA.64.012310}, this was achieved through a Steane-type error correction scheme, which employs two ancilla GKP states $\ket{0}_{\text{GKP}}$ and $\ket{+}_{\text{GKP}}$, the SUM gate and its inverse, along with homodyne measurements. More recently, Ref. \cite{walshe2020continuous} introduced an alternative approach: a teleportation-based GKP error-correction method. This scheme requires two identical ancilla GKP states $\ket{\O}$ (defined below), beam-splitter interactions, and homodyne measurements [see Fig.\ref{fig:Stg123}].
One key advantage of the teleportation-based method is that it relies solely on beam-splitter interactions and does not require any online squeezing operations, as highlighted in Refs. \cite{walshe2020continuous, larsen2021fault, fukui2021all, cohen2022low}. This feature is particularly beneficial in optical implementations, where beam-splitter interactions are significantly easier to realize than squeezing operations. Notably, it has been shown that the teleportation-based method outperforms the Steane-type scheme under identical conditions for the quality of ancilla GKP states. Consequently, the teleportation-based approach is generally preferable to the Steane-type method, regardless of whether online squeezing operations are feasible. Passing two qunaught states through a $50:50$ beam splitter produces a GKP Bell state \cite{walshe2020continuous}.
The beam-splitter unitary operator between modes $j$ and $k$ is defined as

\begin{align}
   \hat{B}_{jk}(\theta) = \exp\big(-i\theta (\hat{q}_j\hat{p}_k - \hat{p}_j\hat{q}_k)\big) 
\end{align}

Applying the balanced beam-splitter interaction  $\hat{B}_{23}(\pi/4)$ between the two GKP qunaught states $\ket{\bar{\O}}_{2}$ and $\ket{\bar{\O}}_{3}$, we get an encoded GKP-Bell state, i.e.,

\begin{align}
    \hat{B}_{23}(\pi/4)\ket{\bar{\O}}_{2}\ket{\bar{\O}}_{3} \equiv \frac{1}{\sqrt{2}}(\ket{0}_{\text{2GKP} }\ket{0}_{3\text{GKP}} + \ket{1}_{2\text{GKP}}\ket{1}_{3\text{GKP}})
\end{align}

Once the GKP-Bell state is prepared, a beam-splitter interaction is applied between modes 1 and 2, followed by homodyne measurements of the quadrature operators $\hat{q}_m$ and $\hat{p}_m$ to correct shift errors $\sqrt{\pi}$ in the case of the GKP code.

In a recent work, a method was introduced for analysing bosonic quantum circuits in the Heisenberg picture, allowing the system’s evolution to be decomposed into distinct signal and noise components. This framework offers clear insight into how loss and noise propagate throughout the circuit \cite{ralph2024noise}.
We begin by specifying the statistical properties of the random displacement errors. The variables $\varsigma_q$ and $\varsigma_p$ are assumed to be independently drawn from zero-mean Gaussian distributions with variance $\sigma^2$. Their expectation values therefore satisfy

\begin{equation}
\langle \varsigma_{2q}^2 \rangle = \sigma^2,
\qquad
\langle \varsigma_{3q}^2 \rangle = \sigma^2,
\end{equation}

\begin{equation}
\langle \varsigma_{2p}^2 \rangle = \sigma^2,
\qquad
\langle \varsigma_{3p}^2 \rangle = \sigma^2.
\end{equation}
A balanced beam splitter mixes the shift errors associated with the two ancilla modes. The resulting quadrature shifts are

\begin{equation}
\varsigma^{+}_{q}
=
\frac{\varsigma_{2q}+\varsigma_{3q}}{\sqrt{2}},
\qquad
\varsigma^{+}_{p}
=
\frac{\varsigma_{2p}+\varsigma_{3p}}{\sqrt{2}},
\end{equation}
\begin{equation}
\varsigma^{-}_{q}
=
\frac{\varsigma_{2q}-\varsigma_{3q}}{\sqrt{2}},
\qquad
\varsigma^{-}_{p}
=
\frac{\varsigma_{2p}-\varsigma_{3p}}{\sqrt{2}},
\end{equation}
To determine the measurement outcomes $q_m$ and $p_m$, 
\begin{align}
\varsigma^{+}_{q} - \varsigma^{-}_{q}
& =
\frac{\varsigma_{2q}+\varsigma_{3q}}{\sqrt{2}}
-
\frac{\varsigma_{2q}-\varsigma_{3q}}{\sqrt{2}}\\
&=
\frac{2\varsigma_{3q}}{\sqrt{2}}
=
\sqrt{2}\,\varsigma_{3q}.
\end{align}
For the $p$ quadrature gives

\begin{equation}
\varsigma^{+}_{p} - \varsigma^{-}_{p}
=
\frac{2\varsigma_{3p}}{\sqrt{2}}
=
\sqrt{2}\,\varsigma_{3p}.
\end{equation}
\begin{equation}
q_m
=
\frac{\varsigma^{+}_{q}-\varsigma^{-}_{q}}{\sqrt{2}}
=
\frac{\sqrt{2}\varsigma_{3q}}{\sqrt{2}}
=
\varsigma_{3q},
\end{equation}

\begin{equation}
p_m
=
\frac{\varsigma^{+}_{p}-\varsigma^{-}_{p}}{\sqrt{2}}
=
\frac{\sqrt{2}\varsigma_{3p}}{\sqrt{2}}
=
\varsigma_{3p}.
\end{equation}
The displacement shifts applied to the data GKP qubit are $\sqrt{2}q_m$ and $\sqrt{2}p_m$. Using the results obtained above, these effective shifts can be written as

\begin{equation}
\sqrt{2}q_m
=
\varsigma^{+}_{q}-\varsigma^{-}_{q}
=
\sqrt{2}\varsigma_{3q},
\end{equation}

\begin{equation}
\sqrt{2}p_m
=
\varsigma^{+}_{p}-\varsigma^{-}_{p}
=
\sqrt{2}\varsigma_{3p}.
\end{equation}

The variances are

\begin{align}
\mathrm{Var}(\sqrt{2}q_m)
&=
\left\langle
\left(
\sqrt{2}\varsigma_{3q}
\right)^2
\right\rangle
\nonumber \\
&=
\left\langle
2\varsigma_{3q}^2
\right\rangle
=
2\langle \varsigma_{3q}^2 \rangle =
2\sigma^2 \\
\mathrm{Var}(\sqrt{2}p_m)
&=
2\sigma^2.
\end{align}
Thus, the logical error probabilities are determined by $
E_{X(Z)}
\left(
2\sigma^2
\right).$
Fig.\ref{ERPROB1} shows the error probability of the GKP qubit after amplification followed by perfect teleportation-based GEC. As expected, the error probability decreases with increasing squeezing. However, perfect teleportation-based GEC is not practically achievable. Therefore, in Fig.\ref{fig:Clipping}, we present the error probabilities as a function of squeezing for imperfect teleportation-based GEC, where the resource states are approximate GKP states and a clipping method is applied. The clipping method reduces the error probability at the expense of a reduced success probability. By looking into the success probabilities shown in Fig.\ref{plot:succ}, we select the dashed deep sky-blue line as representing the optimal clipping.

\subsection{Teleportation-based GKP Error Correction with loss and Amplification}
\label{Teleportation-based GKP Error Correction with loss}

To account for photon loss on the data qubit before entering the teleportation circuit, we model the process as a pure loss channel with transmissivity $\eta$.
In the Heisenberg picture \cite{ralph2024noise}, the loss channel couples the data mode to an environmental vacuum mode ($v_1$), leading to the transformations
\begin{align}
    \hat{q}_{\text{lossy}} = \sqrt{\eta}\hat{q}_{\text{data}} + \sqrt{1-\eta}\hat{q}_{v_1}
\end{align}
To recover the signal amplitude, an amplifier is applied with amplitude gain $g = \frac{1}{\sqrt{\eta}}$.
This amplification introduces to an additional independent vacuum mode $\hat{v}_2$
\begin{align}
    \hat{q}_{\text{amp}} = g\hat{q}_{\text{lossy}} + \sqrt{g^2 - 1}\hat{q}_{v2}
\end{align}
Substituting,
\begin{align}
    \hat{q}_{\text{amp}} & =
\frac{1}{\sqrt{\eta}}
\left(
\sqrt{\eta}\hat{q}_{\text{data}} +
\sqrt{1-\eta}\hat{q}_{v1}
\right)
+ \sqrt{\frac{1}{\eta}-1}\hat{q}_{v2} \\
&=
\hat{q}_{\text{data}} +
\sqrt{\frac{1-\eta}{\eta}}\hat{q}_{v1} +
\sqrt{\frac{1-\eta}{\eta}}\hat{q}_{v2}
\end{align}
Thus, the effective displacement noise on the amplified data qubit contains two statistically independent vacuum noise contributions. Since both are zero-mean Gaussian random variables, their variances sum as
\begin{align}
  \langle  \varsigma^2_{\text{amp}, q}, \varsigma^2_{\text{amp}, p} \rangle
& = 
\frac{1-\eta}{2\eta} + \frac{1-\eta}{2\eta}\\
& = \frac{1-\eta}{\eta}
\end{align}
The ancilla modes (modes 2 and 3) each experience Gaussian displacement noise is $\langle \varsigma_2^2 \rangle = \sigma^2 $ and $\langle \varsigma_3^2 \rangle = \sigma^2 $.
Following balanced beam-splitter interaction, the transformed quadrature shifts are
\begin{align}
    \varsigma'_{+q}  & =
\frac{\varsigma_{2q} + \varsigma_{3q}}{\sqrt{2}} \\
\varsigma'_{-q} & =
\frac{\varsigma_{2q} - \varsigma_{3q}}{\sqrt{2}}
\end{align}
After the beam splitter interaction with data qubit and homodyne measurement,
\begin{align}
    q_m & =
\frac{
\varsigma_{\text{amp}, q}
-
\varsigma'_{-q}
}{
\sqrt{2}
} \\
p_m & =
\frac{
\varsigma_{\text{amp}, p}
-
\varsigma'_{-p}
}{
\sqrt{2}
}
\end{align}
For the $q$ quadrature
\begin{align}
    \varsigma_{\text{total}, q}
& =
\sqrt{2}
\left(
\frac{
\varsigma_{\text{amp}, q}
-
\varsigma'_{-q}
}{
\sqrt{2}
}
\right)
+
\varsigma'_{+q} \\
& = \varsigma_{\text{amp}, q}
+
(\varsigma'_{+q} - \varsigma'_{-q})
\end{align}

Since the loss-induced displacement and ancilla preparation noise are statistically independent, their variances add
\begin{align}
\mathrm{Var}(\varsigma'_{+q}-\varsigma'_{-q})
&
=
2\sigma^2, \\
\text{Var}(\varsigma_{\text{amp}, q})
    & =
    \frac{1-\eta}{\eta}
\end{align}
Therefore, the total variance on the teleported data qubit is $\mathrm{Var}(\varsigma_{\mathrm{total},q})
=
2\sigma^2 + \frac{1-\eta}{\eta}.$
By symmetry, the same expression applies to the $p$-quadrature is $\mathrm{Var}(\varsigma_{\mathrm{total},p})
=
2\sigma^2 + \frac{1-\eta}{\eta}.$
Thus, the overall logical error probability during GKP error correction is determined by $E_{X(Z)}
\left(
2\sigma^2 + \frac{1-\eta}{\eta}
\right).$

\subsection{Teleportation-based GKP Error Correction with telemplification}
\label{Teleportation-based GKP Error Correction with telemplification}
Data Qubit Undergoes phase-insensitive amplification and propagates through a lossy channel with transmissivity $T = \sqrt{\eta}$. The random displacement introduced by this channel is denoted as $\varsigma_{L1}$. The variance for both $q$ and $p$ quadratures is $ \langle \varsigma^2_{L1,q}, \varsigma^2_{L1,p}\rangle = \frac{1-\sqrt{\eta}}{2\sqrt{\eta}}$.
Modes 2 and 3 (Ancilla resource states) have an intrinsic finite-energy GKP squeezing variance, denoted as $\sigma^2$ for each independent mode ($\varsigma_{2q}$ and $\varsigma_{3q}$).
An entangled resource state is prepared by mixing Modes 2 and 3 on a balanced (50:50) beam splitter. The resulting displacements are
\begin{align}
\varsigma'_{+q} &= \frac{\varsigma_{2q} + \varsigma_{3q}}{\sqrt{2}} \\
\varsigma'_{-q} &= \frac{\varsigma_{2q} - \varsigma_{3q}}{\sqrt{2}}
\end{align}

Before entering the teleportation protocol, the ancilla mode ($\varsigma'_{-}$) undergoes a relay transformation with vacuum noise, introducing a new displacement $\varsigma_{L2}$. Its variance is $ \langle \varsigma^2_{L2,q}, \varsigma^2_{L2,p}\rangle = \frac{1-\sqrt{\eta}}{2\sqrt{\eta}}$.
The data mode (carrying $\varsigma_{L1}$) and the ancilla relay mode (carrying $\varsigma'_{-}$ and $\varsigma_{L2}$) are mixed on a balanced beam splitter.
Define a composite channel noise variable, $\varsigma_{(L1,L2)}$, which represents the combined linear combination of the independent channel and relay noises interfering at the beam splitter: $\varsigma_{(L1,L2),q} = \varsigma_{L1,q} - \varsigma_{L2,q}$
Because $\varsigma_{L1}$ and $\varsigma_{L2}$ are independent zero-mean Gaussian variables, their variances add
\begin{align}
\mathrm{Var}(\varsigma_{(L1,L2),q})
&= \mathrm{Var}(\varsigma_{L1,q}) + \mathrm{Var}(\varsigma_{L2,q}) \\
&= \frac{1-\sqrt{\eta}}{2\sqrt{\eta}} + \frac{1-\sqrt{\eta}}{2\sqrt{\eta}} \\
&= \frac{1-\sqrt{\eta}}{\sqrt{\eta}}
\end{align}
After homodyne measurement, the outcomes $q_m$ and $p_m$ capture both the composite channel noise and the intrinsic ancilla noise
\begin{align}
q_m &= \frac{\varsigma_{(L1,L2),q} - \varsigma'_{-q}}{\sqrt{2}} \\
p_m &= \frac{\varsigma_{(L1,L2),p} - \varsigma'_{-p}}{\sqrt{2}}
\end{align}
To complete the teleportation, a feedforward operation displaces the remaining ancilla mode ($\varsigma'_{+}$) by a factor of $\sqrt{2}$ times the measurement outcomes.
For the $q$-quadrature, the final displacement on the teleported GKP data qubit is
\begin{align}
\varsigma_{\mathrm{total},q}
&= \sqrt{2}q_m + \varsigma'_{+q} \\
&= \sqrt{2}\left(\frac{\varsigma_{(L1,L2),q} - \varsigma'_{-q}}{\sqrt{2}}\right) + \varsigma'_{+q} \\
&= \varsigma_{(L1,L2),q} - \varsigma'_{-q} + \varsigma'_{+q} \\
&= \varsigma_{(L1,L2),q} + (\varsigma'_{+q} - \varsigma'_{-q})
\end{align}

To determine the variance of $\varsigma_{\mathrm{total},q}$, we evaluate the two independent contributions separately.
First, for the intrinsic GKP noise term
\begin{align}
\varsigma'_{+q} - \varsigma'_{-q}
&= \left(\frac{\varsigma_{2q} + \varsigma_{3q}}{\sqrt{2}}\right)
 - \left(\frac{\varsigma_{2q} - \varsigma_{3q}}{\sqrt{2}}\right) \\
&= \frac{2\varsigma_{3q}}{\sqrt{2}} \\
&= \sqrt{2}\varsigma_{3q}
\end{align}
Since $\mathrm{Var}(\varsigma_{3q}) = \sigma^2$, we obtain $\mathrm{Var}(\varsigma'_{+q} - \varsigma'_{-q})
= \mathrm{Var}(\sqrt{2}\varsigma_{3q})
= 2\sigma^2$
Second, for the composite channel noise term $\mathrm{Var}(\varsigma_{(L1,L2),q})
= 2\left(\frac{1-\sqrt{\eta}}{\sqrt{\eta}}\right)$
The total variance is $\mathrm{Var}(\varsigma_{\mathrm{total},q})
= 2\sigma^2 + \left(\frac{1-\sqrt{\eta}}{\sqrt{\eta}}\right)$
By symmetry, the same result applies to the $p$-quadrature $\mathrm{Var}(\varsigma_{\mathrm{total},p})
= 2\sigma^2 + \left(\frac{1-\sqrt{\eta}}{\sqrt{\eta}}\right)$
Thus, the logical error probabilities are determined by $
E_{X(Z)}
\left(
2\sigma^2
+
\frac{1-\sqrt{\eta}}{\sqrt{\eta}}
\right).$

\section{More details on GKP-Parity-Encoding}
\label{More details on GKP-Parity-Encoding}
\subsection{Bell-state Measurement of GKP qubits}
The computational basis states of the ideal GKP qubit are given by
\begin{align}
    \ket{\bar{0}}_{\text{GKP}} & \equiv \sum_{n \in \mathbb{Z}} \ket{\hat{q} = 2n\sqrt{\pi}} \\
    \ket{\bar{1}}_{\text{GKP}} & \equiv \sum_{n \in \mathbb{Z}} \ket{\hat{q} = (2n + 1)\sqrt{\pi}} \\
    \ket{\bar{+}}_{\text{GKP}} & \equiv \sum_{n \in \mathbb{Z}} \ket{\hat{p} = 2n\sqrt{\pi}} \\
    \ket{\bar{-}}_{\text{GKP}} & \equiv \sum_{n \in \mathbb{Z}} \ket{\hat{p} = (2n + 1)\sqrt{\pi}}
\end{align}
\begin{align}
    \ket{\bar{0}}_{\text{GKP}} & \equiv \sum_{n \in \mathbb{Z}} \ket{\hat{q} = 2n\sqrt{\pi}} \\
   & \equiv \sum_{n \in  \mathbb{Z}} \delta(q - 2n \sqrt{\pi})\ket{q}   \\
   & \equiv  \int \,dp \frac{1}{\sqrt{2 \pi}} \sum_{n \in  \mathbb{Z}} e^{i p 2n \sqrt{\pi}} \ket{p} \\
  & \equiv  \frac{1}{\sqrt{2}} \sum_{n \in  \mathbb{Z}}   \delta(p - n \sqrt{\pi})\ket{p} \\
  & \equiv  \frac{1}{\sqrt{2}} (\ket{\bar{+}}_{\text{GKP}} + \ket{\bar{-}}_{\text{GKP}})
\end{align}
Similarly, 
\begin{align}
    \ket{\bar{1}}_{\text{GKP}} & \equiv  \frac{1}{\sqrt{2}} (\ket{\bar{+}}_{\text{GKP}} - \ket{\bar{-}}_{\text{GKP}})
\end{align}
The four Bell states of GKP qubits are
\begin{align}
    \ket{\phi}_{\pm} & \equiv \ket{\bar{0}}_{\text{GKP}}\ket{\bar{0}}_{\text{GKP}} \pm \ket{\bar{1}}_{\text{GKP}}\ket{\bar{1}}_{\text{GKP}} \\
    \ket{\psi}_{\pm} & \equiv \ket{\bar{0}}_{\text{GKP}}\ket{\bar{1}}_{\text{GKP}} \pm \ket{\bar{1}}_{\text{GKP}}\ket{\bar{0}}_{\text{GKP}}
\end{align}

A Bell-state measurement (BSM) of lossless GKP qubits is implemented using a 50:50 beam splitter followed by homodyne measurement, as illustrated in Fig.~\ref{fig:CBSMmain}. Except for the case homodyne fails, the four Bell states can be deterministically identified from the detector outcomes. Specifically, the measurement results correspond to the Bell states as follows:$ (\text{even}, 0) \to \ket{\phi}_{+},\: (\text{odd}, 0)  \to \ket{\phi}_{-}$, $ ( 0, \text{even})  \to \ket{\psi}_{+}, \:  ( 0, \text{odd})  \to \ket{\psi}_{-}$.  If the homodyne readout falls significantly outside the regions of the lattice, the event is discarded as a BSM failure. In this case, only the sign of the Bell state (i.e., the $\pm$ in $\ket{\phi}_{\pm}$ or $\ket{\psi}_{\pm}$) can be determined, since there is an intrinsic ambiguity between $\ket{\phi}_{+}$ and $\ket{\psi}_{+}$. For realistic scenarios, photon loss and finite squeezing effects must be taken into account. Photon loss can be modelled by a beamsplitter interaction in which each mode is independently mixed with the vacuum state. The final state after photon loss is obtained by tracing out the vacuum modes from the output. Finally, the probability of misidentifying the bit value of an approximate GKP qubit, denoted by $P_{\text{Fail}}(\sigma^2)$, is approximately given by $P_{\text{Fail}}(\sigma^2) = 1-  \int^{\sqrt{\pi}/2}_{-\sqrt{\pi}/2} \,dx \frac{1}{\sqrt{2 \pi \sigma^{2} }} e^{-x^2/ 2\sigma^{2}} $.

When the state prior to photon loss is one of the four Bell states, the measurement outcomes can be classified into: success, $X$-error, $Z$-error, $Y$-error, or failure. If the final Bell state coincides with the initial one, the outcome is regarded as a success.
An $X$-error corresponds to a letter flip, namely a change in the Bell-state label ($\phi$ or $\psi$), for example from $\ket{\phi}_{+}$ to $\ket{\psi}_{+}$. A $Z$-error corresponds to a sign flip, i.e., a change in the relative phase ($\pm$) of the Bell state, such as from $\ket{\phi}_{+}$ to $\ket{\phi}_{-}$. A $Y$-error represents the simultaneous occurrence of both a letter flip and a sign flip. The final case, referred to as a failure, corresponds to the measurement outcomes $(0,0)$, $(\text{even},\text{even})$, and $(\text{odd},\text{odd})$, for which the Bell-state letter cannot be uniquely identified.

\subsection{Decomposition of the Encoded Bell States}
\label{Decomposition of the Encoded Bell States}
In the GKP-parity-state-encoding,
\begin{align}
    \ket{0}_{\text{L}} & \equiv [\ket{\bar{+}}^{\otimes m}_{\text{GKP}} + \ket{\bar{-}}^{\otimes m}_{\text{GKP}}]_{1} [\ket{\bar{+}}^{\otimes m}_{\text{GKP}} + \ket{\bar{-}}^{\otimes m}_{\text{GKP}}]_{2} \dots [\ket{\bar{+}}^{\otimes m}_{\text{GKP}} + \ket{\bar{-}}^{\otimes m}_{\text{GKP}}]_{n} \\
     \ket{1}_{\text{L}}   & \equiv [\ket{\bar{+}}^{\otimes m}_{\text{GKP}} - \ket{\bar{-}}^{\otimes m}_{\text{GKP}}]_{1} [\ket{\bar{+}}^{\otimes m}_{\text{GKP}} - \ket{\bar{-}}^{\otimes m}_{\text{GKP}}]_{2} \dots [\ket{\bar{+}}^{\otimes m}_{\text{GKP}} - \ket{\bar{-}}^{\otimes m}_{\text{GKP}}]_{n}
\end{align}
The Bell-states are written as
\begin{align}
    \ket{\phi}_{\pm} & \equiv \frac{1}{\sqrt{2}}\Big[\Big([\ket{\bar{+}}^{\otimes m}_{\text{GKP}} + \ket{\bar{-}}^{\otimes m}_{\text{GKP}}]_{1} \dots [\ket{\bar{+}}^{\otimes m}_{\text{GKP}} + \ket{\bar{-}}^{\otimes m}_{\text{GKP}}]_{n}\Big)\Big([\ket{\bar{+}}^{\otimes m}_{\text{GKP}} + \ket{\bar{-}}^{\otimes m}_{\text{GKP}}]_{1'} \dots [\ket{\bar{+}}^{\otimes m}_{\text{GKP}} + \ket{\bar{-}}^{\otimes m}_{\text{GKP}}]_{n'}\Big) \nonumber
    \\
   & \pm  \Big([\ket{\bar{+}}^{\otimes m}_{\text{GKP}} - \ket{\bar{-}}^{\otimes m}_{\text{GKP}}]_{1}  \dots [\ket{\bar{+}}^{\otimes m}_{\text{GKP}} - \ket{\bar{-}}^{\otimes m}_{\text{GKP}}]_{n}\Big)\Big([\ket{\bar{+}}^{\otimes m}_{\text{GKP}} - \ket{\bar{-}}^{\otimes m}_{\text{GKP}}]_{1'}  \dots [\ket{\bar{+}}^{\otimes m}_{\text{GKP}} - \ket{\bar{-}}^{\otimes m}_{\text{GKP}}]_{n'}\Big)\Big]
\end{align}
\begin{align}
    \ket{\psi}_{\pm}  & \equiv \frac{1}{\sqrt{2}}\Big[ \Big([\ket{\bar{+}}^{\otimes m}_{\text{GKP}} + \ket{\bar{-}}^{\otimes m}_{\text{GKP}}]_{1} \dots [\ket{\bar{+}}^{\otimes m}_{\text{GKP}} + \ket{\bar{-}}^{\otimes m}_{\text{GKP}}]_{n}\Big)\Big([\ket{\bar{+}}^{\otimes m}_{\text{GKP}} + \ket{\bar{-}}^{\otimes m}_{\text{GKP}}]_{1'} \dots [\ket{\bar{+}}^{\otimes m}_{\text{GKP}} + \ket{\bar{-}}^{\otimes m}_{\text{GKP}}]_{n'}\Big) \nonumber \\
   & \pm \Big([\ket{\bar{+}}^{\otimes m}_{\text{GKP}} - \ket{\bar{-}}^{\otimes m}_{\text{GKP}}]_{1}  \dots [\ket{\bar{+}}^{\otimes m}_{\text{GKP}} - \ket{\bar{-}}^{\otimes m}_{\text{GKP}}]_{n}\Big) \Big([\ket{\bar{+}}^{\otimes m}_{\text{GKP}} + \ket{\bar{-}}^{\otimes m}_{\text{GKP}}]_{1'} \dots [\ket{\bar{+}}^{\otimes m}_{\text{GKP}} + \ket{\bar{-}}^{\otimes m}_{\text{GKP}}]_{n'}\Big)  
    \Big]
\end{align}
where the first $n$ blocks correspond to the first qubit, while the subsequent $n'$ blocks are from the second qubit. By rearranging the block ordering from $(1,\dots,n,1',\dots,n')$ to $(1,1',2,2',\dots,n,n')$, the state can be fully decomposed into Bell states at the block level.

\begin{align}
    \ket{\phi}_{+} & \equiv \frac{1}{\sqrt{2}}\Big[\Big([\ket{\bar{+}}^{\otimes m}_{\text{GKP}} + \ket{\bar{-}}^{\otimes m}_{\text{GKP}}]_{1} \dots [\ket{\bar{+}}^{\otimes m}_{\text{GKP}} + \ket{\bar{-}}^{\otimes m}_{\text{GKP}}]_{n}\Big)\Big([\ket{\bar{+}}^{\otimes m}_{\text{GKP}} + \ket{\bar{-}}^{\otimes m}_{\text{GKP}}]_{1'} \dots [\ket{\bar{+}}^{\otimes m}_{\text{GKP}} + \ket{\bar{-}}^{\otimes m}_{\text{GKP}}]_{n'}\Big) \nonumber
    \\
   & +  \Big([\ket{\bar{+}}^{\otimes m}_{\text{GKP}} - \ket{\bar{-}}^{\otimes m}_{\text{GKP}}]_{1}  \dots [\ket{\bar{+}}^{\otimes m}_{\text{GKP}} - \ket{\bar{-}}^{\otimes m}_{\text{GKP}}]_{n}\Big)\Big([\ket{\bar{+}}^{\otimes m}_{\text{GKP}} - \ket{\bar{-}}^{\otimes m}_{\text{GKP}}]_{1'}  \dots [\ket{\bar{+}}^{\otimes m}_{\text{GKP}} - \ket{\bar{-}}^{\otimes m}_{\text{GKP}}]_{n'}\Big)\Big] \\
   & \equiv \frac{1}{\sqrt{2}}\Big[ [\ket{\bar{+}}^{\otimes m}_{\text{GKP}} + \ket{\bar{-}}^{\otimes m}_{\text{GKP}}]_{1}[\ket{\bar{+}}^{\otimes m}_{\text{GKP}} + \ket{\bar{-}}^{\otimes m}_{\text{GKP}}]_{1'} \dots [\ket{\bar{+}}^{\otimes m}_{\text{GKP}} + \ket{\bar{-}}^{\otimes m}_{\text{GKP}}]_{n}[\ket{\bar{+}}^{\otimes m}_{\text{GKP}} + \ket{\bar{-}}^{\otimes m}_{\text{GKP}}]_{n'} \nonumber \\
   & + [\ket{\bar{+}}^{\otimes m}_{\text{GKP}} - \ket{\bar{-}}^{\otimes m}_{\text{GKP}}]_{1}[\ket{\bar{+}}^{\otimes m}_{\text{GKP}} - \ket{\bar{-}}^{\otimes m}_{\text{GKP}}]_{1'} \dots [\ket{\bar{+}}^{\otimes m}_{\text{GKP}} - \ket{\bar{-}}^{\otimes m}_{\text{GKP}}]_{n} [\ket{\bar{+}}^{\otimes m}_{\text{GKP}} - \ket{\bar{-}}^{\otimes m}_{\text{GKP}}]_{n'}  
   \Big] \\
 & \equiv  \frac{1}{\sqrt{2^{n-1}}}\Big(\ket{\phi^{m}_{+}}_{11'} \ket{\phi^{m}_{+}}_{22'}\dots \ket{\phi^{m}_{+}}_{nn'} + \ket{\phi^{m}_{-}}_{11'}\ket{\phi^{m}_{-}}_{22'}\ket{\phi^{m}_{+}}_{33'} \dots \ket{\phi^{m}_{+}}_{nn'} \nonumber \\
 & +  \ket{\phi^{m}_{-}}_{11'}\ket{\phi^{m}_{+}}_{22'}\ket{\phi^{m}_{-}}_{33'} \dots \ket{\phi^{m}_{+}}_{nn'} \nonumber \\
 &\vdots \nonumber\\
  &\ket{\phi^{m}_{+}}_{11'}\ket{\phi^{m}_{+}}_{22'}\ket{\phi^{m}_{+}}_{33'} \dots \ket{\phi^{m}_{+}}_{n-2 n-2'}\ket{\phi^{m}_{-}}_{n-1 n-1'} \ket{\phi^{m}_{-}}_{n n'} + \dots\Big) \\
  & \equiv  \frac{1}{\sqrt{2^{n -1}}}\sum_{k = \text{even} \leq n} \mathcal{P}\Big[\ket{\phi^{m}}_{-}^{\otimes k}\ket{\phi^{m}}_{+}^{\otimes n-k}\Big]
\end{align}
which is the equally weighted superposition of all possible $n-$fold tensor products of even number $k$ of $ \ket{\phi^{m}}_{-}$ and $n-k$  of $ \ket{\phi^{m}}_{+}$. Likewise for others, all the (logical) second-level Bell states can be represented by

\begin{align}
 \ket{\phi}_{+(-)} & \equiv \frac{1}{\sqrt{2^{n -1}}}\sum_{k = \text{even(odd)} \leq n} \mathcal{P}\Big[\ket{\phi^{m}}_{-}^{\otimes k}\ket{\phi^{m}}_{+}^{\otimes n-k}\Big] \\
 \ket{\psi}_{+(-)} & \equiv \frac{1}{\sqrt{2^{n -1}}}\sum_{k = \text{even(odd)} \leq n} \mathcal{P}\Big[\ket{\psi^{m}}_{-}^{\otimes k}\ket{\psi^{m}}_{+}^{\otimes n-k}\Big] 
\end{align}

where $\mathcal{P}[.]$ is defined as a permutation function, e.g., $\mathcal{P}[\ket{\phi}_{+}\ket{\phi}_{-}\ket{\phi}_{-}] = \ket{\phi}_{+}\ket{\phi}_{-}\ket{\phi}_{-} + \ket{\phi}_{-}\ket{\phi}_{+}\ket{\phi}_{-} + \ket{\phi}_{-}\ket{\phi}_{-}\ket{\phi}_{+}$.
Now similarly in the physical level Bell states are

\begin{align}
    \ket{\phi^{m}}_{+} & = \frac{1}{\sqrt{2}}\Big(\ket{\bar{+}}^{\text{GKP}}_{1} \dots \ket{\bar{+}}^{\text{GKP}}_{m}\ket{\bar{+}}^{\text{GKP}}_{1'} \dots \ket{\bar{+}}^{\text{GKP}}_{m'} + \ket{\bar{-}}^{\text{GKP}}_{1} \dots \ket{\bar{-}}^{\text{GKP}}_{m}\ket{\bar{-}}^{\text{GKP}}_{1'} \dots \ket{\bar{-}}^{\text{GKP}}_{m'}       \Big) \\
     \ket{\phi^{m}}_{-} & = \frac{1}{\sqrt{2}}\Big(\ket{\bar{+}}^{\text{GKP}}_{1} \dots \ket{\bar{+}}^{\text{GKP}}_{m}\ket{\bar{-}}^{\text{GKP}}_{1'} \dots \ket{\bar{-}}^{\text{GKP}}_{m'} + \ket{\bar{-}}^{\text{GKP}}_{1} \dots \ket{\bar{-}}^{\text{GKP}}_{m}\ket{\bar{+}}^{\text{GKP}}_{1'} \dots \ket{\bar{+}}^{\text{GKP}}_{m'}       \Big) \\
 \ket{\psi^{m}}_{+} & = \frac{1}{\sqrt{2}}\Big(\ket{\bar{+}}^{\text{GKP}}_{1} \dots \ket{\bar{+}}^{\text{GKP}}_{m}\ket{\bar{+}}^{\text{GKP}}_{1'} \dots \ket{\bar{+}}^{\text{GKP}}_{m'} - \ket{\bar{-}}^{\text{GKP}}_{1} \dots \ket{\bar{-}}^{\text{GKP}}_{m}\ket{\bar{-}}^{\text{GKP}}_{1'} \dots \ket{\bar{-}}^{\text{GKP}}_{m'}\\
  \ket{\psi^{m}}_{-} & = \frac{1}{\sqrt{2}}\Big(\ket{\bar{+}}^{\text{GKP}}_{1} \dots \ket{\bar{+}}^{\text{GKP}}_{m}\ket{\bar{-}}^{\text{GKP}}_{1'} \dots \ket{\bar{-}}^{\text{GKP}}_{m'} - \ket{\bar{-}}^{\text{GKP}}_{1} \dots \ket{\bar{-}}^{\text{GKP}}_{m}\ket{\bar{+}}^{\text{GKP}}_{1'} \dots \ket{\bar{+}}^{\text{GKP}}_{m'}       \Big)
\end{align}
Now by rearranging the order to $(1, 1',2,2',\dots m, m')$
\begin{align}
    \ket{\phi^{m}}_{+} & = \frac{1}{\sqrt{2}}\Big(\ket{\bar{+}}^{\text{GKP}}_{1} \dots \ket{\bar{+}}^{\text{GKP}}_{m}\ket{\bar{+}}^{\text{GKP}}_{1'} \dots \ket{\bar{+}}^{\text{GKP}}_{m'} + \ket{\bar{-}}^{\text{GKP}}_{1} \dots \ket{\bar{-}}^{\text{GKP}}_{m}\ket{\bar{-}}^{\text{GKP}}_{1'} \dots \ket{\bar{-}}^{\text{GKP}}_{m'}       \Big)  \\
   & =   \frac{1}{\sqrt{2}}\Big( \ket{\bar{+}}^{\text{GKP}}_{1} \ket{\bar{+}}^{\text{GKP}}_{1'} \ket{\bar{+}}^{\text{GKP}}_{2} \ket{\bar{+}}^{\text{GKP}}_{2'}  \dots \ket{\bar{+}}^{\text{GKP}}_{m} \ket{\bar{+}}^{\text{GKP}}_{m'} + \ket{\bar{-}}^{\text{GKP}}_{1} \ket{\bar{-}}^{\text{GKP}}_{1'} \ket{\bar{-}}^{\text{GKP}}_{2} \ket{\bar{-}}^{\text{GKP}}_{2'}  \dots \ket{\bar{-}}^{\text{GKP}}_{m} \ket{\bar{-}}^{\text{GKP}}_{m'} \Big) \\
   & =  \frac{1}{\sqrt{2^{n-1}}}\Big(\ket{\phi_{+}}_{11'} \ket{\phi_{+}}_{22'}\dots \ket{\phi_{+}}_{mm'} + \ket{\phi_{-}}_{11'}\ket{\phi_{-}}_{22'}\ket{\phi_{+}}_{33'} \dots \ket{\phi_{+}}_{mm'} \nonumber \\
 & +  \ket{\phi_{-}}_{11'}\ket{\phi_{+}}_{22'}\ket{\phi_{-}}_{33'} \dots \ket{\phi_{+}}_{mm'} \nonumber \\
 &\vdots \nonumber\\
  &\ket{\phi_{+}}_{11'}\ket{\phi_{+}}_{22'}\ket{\phi_{+}}_{33'} \dots \ket{\phi_{+}}_{m-2 m-2'}\ket{\phi_{-}}_{m-1 m-1'} \ket{\phi_{-}}_{m m'} + \dots\Big)  \\
  & \equiv \frac{1}{\sqrt{2}} \sum_{l = \text{even}\leq m} \mathcal{P}\Big[\ket{\psi}_{+}^{\otimes l}\ket{\phi}_{+}^{\otimes m-l}\Big]
\end{align}
in the form of the equally weighted superposition of all possible m-fold tensor products of even number $l$ of $\ket{\psi}_{+}$ and $m-l$ of $\ket{\phi_{+}}$. Likewise for others, we can rewrite all the physical-level Bell states as
\begin{align}
     \ket{\phi^{m}}_{\pm} & \equiv \frac{1}{\sqrt{2}} \sum_{l = \text{even}\leq m} \mathcal{P}\Big[\ket{\psi}_{\pm}^{\otimes l}\ket{\phi}_{\pm}^{\otimes m-l}\Big] \\
    \ket{\psi^{m}}_{\pm} & \equiv \frac{1}{\sqrt{2}} \sum_{l = \text{odd}\leq m} \mathcal{P}\Big[\ket{\psi}_{\pm}^{\otimes l}\ket{\phi}_{\pm}^{\otimes m-l}\Big] 
\end{align}

\subsection{Simulation details of the GKP-parity-encoding}
\label{Simulation details of the GKP-parity-encoding}
By imposing a clipping threshold, the clipping protocol evaluates the analog homodyne outcome and discards measurements that fall within an unreliable region. In each block of the protocol, CBSM is performed on two GKP qubits. One of the qubits propagates between neighbouring repeater nodes over a distance $L_{0}$, while the other remains stationary within the repeater. The success probability of this operation is denoted by $P(\eta_{L_{0}}, \eta_{0})$ where $\eta_{0}$ represents the effective loss rate inside the repeater and $\eta_{L_{0}}$ is the effective loss rate of the travelling qubit over the distance $L_{0}$. These two parameters are related through $\eta_{L_{0}} = \eta_{0} e^{-L_0/ L_{\text{att}}}$ with the attenuation length $L_{\text{att}} = 22$km.
At the logical level, the Bell-state measurement allows discrimination between $\ket{\phi}_{\pm}$ and $\ket{\psi}_{\pm}$ even in the presence of photon loss and finite squeezing. Importantly, photon losses occurring in any block-level BSM do not affect the outcomes of the other block-level BSMs. Therefore, by collecting the outcomes from the $n$ independent block-level BSMs, one can fully discriminate $\ket{\phi}_{\pm}$ and $\ket{\psi}_{\pm}$ unless at least one block-level BSM fails.
Using this clipping framework, we derive the complete logical success probability $P_s(\eta_{0},\eta_{L_{0}})$.

Transmission losses $\eta_{0}$ and $\eta_{L_{0}}$ contribute to the total Gaussian noise variance:

\begin{equation}
\sigma_{\mathrm{tot}}^2(\eta_{0},\eta_{L_{0}}) =
2\sigma_{\mathrm{GKP}}^2
+ \frac{1-\eta_{0}}{\eta_{0}}
+ \frac{1-\eta_{L_{0}}}{\eta_{L_{0}}}.
\end{equation}

For GKP qubits, the canonical quadratures are discretized into bins of width $\sqrt{\pi}$. Measurement outcomes $v$ are therefore assigned according to these periodic bins. To suppress unreliable outcomes, a clipping threshold $\delta$ is imposed around each decision boundary at $v = (k \pm 1/2)\sqrt{\pi}$ where measurements falling within the threshold region are discarded. The physical-level outcome probabilities are parameterised with variance $\sigma_{\mathrm{tot}}^2$: (i) correct measurement probability ($P^{(c)}(\sigma_{\mathrm{tot}}^2)$) corresponding to integration over reliable correct bins. (ii) incorrect measurement probability ($P^{(i)}(\sigma_{\mathrm{tot}}^2)$) corresponding to integration over reliable shifted bins.  (iii) discard (erasure) probability ($P^{(d)}(\sigma_{\mathrm{tot}}^2)$) corresponding to integration over threshold regions. For the $j$th physical measurement in block $i$, we assign $m_{p(i,j)} \in \{+1,-1,0\},$ representing correct, incorrect, and discarded outcomes, respectively.
Each block consists of $m$ physical qubits. In the X basis, the block-level parity outcome is defined as $M_{p_i}^{(b)} =
\prod_{j=1}^{m} m_{p(i,j)}.$
If any physical measurement is discarded, i.e., $m_{p(i,j)}=0$ for at least one $j$, then the entire block is treated as erased ($M_{p_i}^{(b)} = 0$). Thus, local clipping events propagate directly into block-level erasures.

A block survives only if all $m$ constituent physical qubits survive. Since a single physical qubit is retained with probability $1 - P^{(d)}(\sigma_{\mathrm{tot}}^2),$
the probability that the entire block survives is $(1 - P^{(d)}(\sigma_{\mathrm{tot}}^2))^m$.
Therefore, the probability that the entire block is discarded is
\begin{align}
    P_{blkX}^{(d)}(\sigma_{\mathrm{tot}}^2)
= 1 - (1 - P^{(d)}(\sigma_{\mathrm{tot}}^2))^m.
\end{align}
\begin{equation}
P_{blkX}^{(c)}(\sigma_{\mathrm{tot}}^2)
=
\sum_{j=0}^{\lfloor m/2 \rfloor}
\binom{m}{2j}
\left(P^{(i)}(\sigma_{\mathrm{tot}}^2)\right)^{2j}
\left(P^{(c)}(\sigma_{\mathrm{tot}}^2)\right)^{m-2j}.
\end{equation}
Using,

\begin{equation}
(x+y)^m + (x-y)^m
=
2
\sum_{j}
\binom{m}{2j}
x^{m-2j}y^{2j},
\end{equation}
\begin{equation}
P_{blkX}^{(c)}(\sigma_{\mathrm{tot}}^2)
=
\frac{
\left(P^{(c)}(\sigma_{\mathrm{tot}}^2) + P^{(i)}(\sigma_{\mathrm{tot}}^2)\right)^m
+
\left(P^{(c)}(\sigma_{\mathrm{tot}}^2) - P^{(i)}(\sigma_{\mathrm{tot}}^2)\right)^m
}{2}.
\end{equation}

Since $P^{(c)}(\sigma_{\mathrm{tot}}^2) + P^{(i)}(\sigma_{\mathrm{tot}}^2)
=
1 - P^{(d)}(\sigma_{\mathrm{tot}}^2),$
we obtain
\begin{equation}
P_{blkX}^{(c)}(\sigma_{\mathrm{tot}}^2)
=
\frac{
(1 - P^{(d)}(\sigma_{\mathrm{tot}}^2))^m
+
(P^{(c)}(\sigma_{\mathrm{tot}}^2) - P^{(i)}(\sigma_{\mathrm{tot}}^2))^m
}{2}.
\end{equation}
The probability of an incorrect block level parity is
\begin{equation}
P_{blkX}^{(i)}(\sigma_{\mathrm{tot}}^2)
=
1
-
P_{blkX}^{(c)}(\sigma_{\mathrm{tot}}^2)
-
P_{blkX}^{(d)}(\sigma_{\mathrm{tot}}^2).
\end{equation}

To determine the final logical bit value, a majority vote is performed across the $n$ block-level measurement outcomes. If $n_d$ blocks are discarded, then the decoder performs majority voting over the remaining $n' = n - n_d$. The full logical success probability is obtained by summing over all possible block outcome combinations, weighted according to their voting success:

\begin{align}
P(\eta_{L_0}, \eta_0) 
&=
\sum_{n_c=0}^{n}
\sum_{n_e=0}^{n-n_c}
\left(
\frac{n!}{n_c! n_e! (n-n_c-n_e)!}
P_s^{n_c}
P_e^{n_e}
p_f^{\,n-n_c-n_e}
\right)
I(n_c,n_e),\\
&= \sum_{n_c=0}^n \sum_{n_e=0}^{n-n_c} \frac{n!}{n_c! n_e! n_d!} \left[ \frac{(1 - P^{(d)})^m + (P^{(c)} - P^{(i)})^m}{2} \right]^{n_c} \nonumber\\
&\left[ \frac{(1 - P^{(d)})^m - (P^{(c)} - P^{(i)})^m}{2} \right]^{n_e} \left[ 1 - (1 - P^{(d)})^m \right]^{n_d} \times I(n_c, n_e)
\end{align}
where the indicator function $I(n_c,n_e)$ is defined as
\begin{equation}
I(n_c,n_e)
=
\begin{cases}
1, & n_c > n_e, \\
\frac{1}{2}, & n_c = n_e, \\
0, & n_c < n_e.
\end{cases}
\end{equation}

\section{Alternative way of GKP-parity-encoding}
\label{Alternative way of GKP-parity-encoding}
To optimize the transmission of a logical qubit over a distance $L$,  we perform numerical searches to determine the optimal parameters. The optimized parameters, denoted as $RC(n, m, j, L_{0})$, are chosen to minimize the total qubit cost $RC$, while accounting for channel losses and finite-squeezing effects that affect GKP qubits during both transmission and repeater operations.
While analysing the resource cost of the scheme, we consider $(\eta_{0} = 0.98)$ and $(\eta_{0} = 0.93)$, and select the following parameter ranges for GKP-parity encoding with CV Bell measurement for the scheme described in Fig. \ref{fig:CBSMmain}. For $(\eta_{0} = 0.98)$, $j$ is chosen from $(j \in {0,1,2})$, with $(m \in [5, 8])$ and $(n \in [15,29])$. $L_{0}$ is  $(1.0\text{km} \leq L_{0} \leq 1.8\text{km})$.
For $(\eta_{0} = 0.93)$, $j$ is again taken from $(j \in {0,1,2})$, while $m$ is varied over $(6 \leq m \leq 11)$ and $n$ over $(45 \leq n \leq 90)$. In this case, $L_{0}$ is chosen within the same interval $(1.0\text{km} \leq L_{0} \leq 1.8\text{km})$.

\begin{figure}[!htb]
	\centering
		 \includegraphics[width=0.5\linewidth]{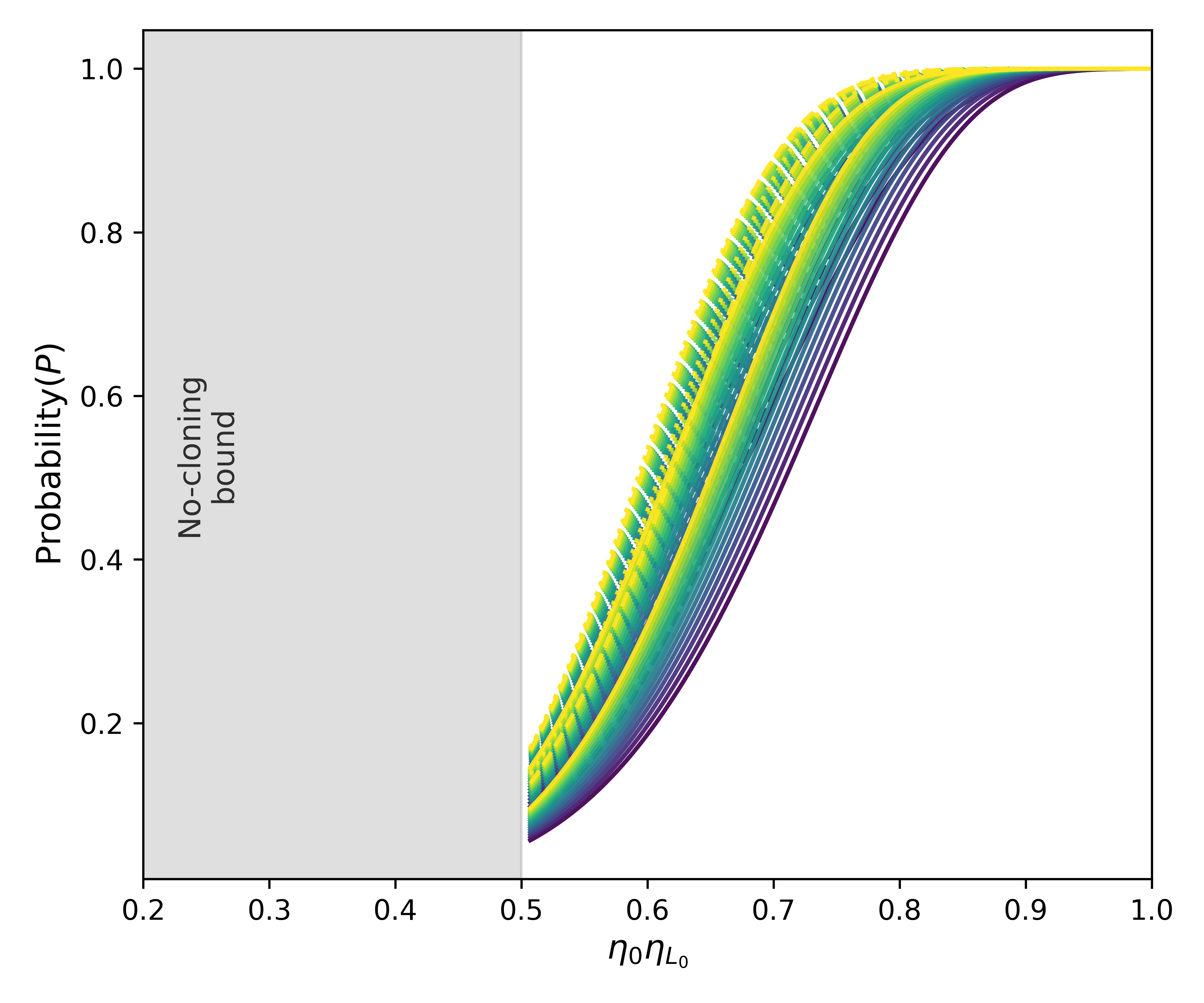}
		\caption{ 
        Success probability as a function of the effective transmissivity $\eta_{0}\eta_{L_{0}}$. Each corresponds to a different combination of $n$, $j$, $m$, and initial loss parameter $L_0$ (sampled uniformly in the interval $[L_{0,\min} (1 \text{km}),\, L_{0,\max}(1.8 \text{km})]$) while keeping the squeezing constant $r = 15.0$dB. The shaded gray region indicates the no-cloning bound.
  \label{noclone}}
\end{figure}

Having discussed GKP-parity encoding using homodyne measurement with clipping (Fig.~\ref{fig:CBSMmain}), we now explore an alternative implementation utilizing two photon number resolving detectors (PNRDs). This approach parallels methods previously applied to photonic and coherent state qubits~\cite{lee2018fundamental, lee2021loss}.

We define the physical-level Bell-state measurement (BSM) for GKP Bell pairs using a standard linear-optical scheme comprising a beam splitter and photon detections. This configuration allows for the deterministic identification of two specific Bell states, $ \ket{\phi}_{\pm}$ and $ \ket{\psi}_{\pm}$. Events where neither detector registers a photon are treated as BSM failures. Structurally, each logical-level BSM is composed of n block-level BSMs, with each block-level BSM further realized by m physical-level BSMs, similar to Fig.~\ref{fig:CBSMmain}. A limitation of using two PNRDs is that the loss tolerance is fundamentally bounded by the condition $\eta_{0}\eta_{L_{0}} > 0.5 $. This restriction arises from the no-cloning theorem, as discussed in prior literature~\cite{muralidharan2014ultrafast, varnava2006loss} and demonstrated for general joint measurements under photon loss by Lee et al.~\cite{lee2018fundamental}. We illustrate this constraint specifically for GKP qubits in Fig.~\ref{noclone}.

Our resource cost optimisation analysis indicates that GKP-parity encoding with PNRDs necessitates a high squeezing parameter to be effective. However, even with high squeezing, the optimized resource cost remains higher than that of the homodyne-with-clipping approach. This performance gap stems from the clipping in homodyne detection, which effectively confines the GKP spikes within the domain, thereby reducing shift errors. Consequently, GKP-parity encoding with homodyne detection proves to be not only optimal in terms of resource cost but also more practically feasible, given the high cost and limited availability of PNRDs in standard experimental setups.

In the following results, we examine various transmission distances while maintaining a constant squeezing parameter of r= 15.0dB. 
In the following, we also denote the logical error by $E_{X(Z)}$.

(i) for 1000 KM,\\

$\eta_{0} = 0.98$, $RC_{\text{min}}  = 1.599 \times 10^5$, rate = 0.7920, j = 1, m = 6, n = 19, $L_{0} = 1.80$ KM, $ E_{X(Z)} =7.365 \times 10^{-6} $.\\

$\eta_{0} = 0.93$, $RC_{\text{min}}  = 8.763 \times 10^5$, rate = 0.6573, j = 1, m = 8, n = 61, $L_{0} = 1.69$ KM, $ E_{X(Z)} =9.819 \times 10^{-6} $.\\

(ii) for 5000 KM,\\
$\eta_{0} = 0.98$, $RC_{\text{min}}  = 1.242 \times 10^6$, rate = 0.7406, j = 2, m = 7, n = 19, $L_{0} = 1.45$ KM, $ E_{X(Z)} =8.582 \times 10^{-6} $.\\

$\eta_{0} = 0.93$, $RC_{\text{min}}  = 8.164 \times 10^6$, rate = 0.6658, j = 1, m = 9, n = 84, $L_{0} = 1.39$ KM, $ E_{X(Z)} =1.105 \times 10^{-5} $.\\

(iii) for 10000 KM,\\

$\eta_{0} = 0.98$, $RC_{\text{min}}  = 2.954 \times 10^6$, rate = 0.7819, j = 1, m = 7, n = 27, $L_{0} = 1.66$ KM, $ E_{X(Z)} =8.582 \times 10^{-6}  $. \\

$\eta_{0} = 0.93$, $RC_{\text{min}}  = 2.083 \times 10^7$, rate = 0.6874, j = 2, m = 10, n = 89, $L_{0} = 1.24$ KM, $ E_{X(Z)} =1.227 \times 10^{-5} $.\\

\end{widetext}

\bibliography{references}

@article{lee2021loss,
  title={Loss-tolerant concatenated Bell-state measurement with encoded coherent-state qubits for long-range quantum communication},
  author={Lee, Seok-Hyung and Lee, Seung-Woo and Jeong, Hyunseok},
  journal={Physical Review Research},
  volume={3},
  number={4},
  pages={043205},
  year={2021},
  publisher={APS}
}

@article{lee2018fundamental,
  title={Fundamental building block for all-optical scalable quantum networks},
  author={Lee, Seung-Woo and Ralph, Timothy C and Jeong, Hyunseok},
  journal={arXiv preprint arXiv:1804.09342},
  year={2018}
}

@article{ralph2005loss,
  title={Loss-tolerant optical qubits},
  author={Ralph, Timothy C and Hayes, AJF and Gilchrist, Alexei},
  journal={Physical review letters},
  volume={95},
  number={10},
  pages={100501},
  year={2005},
  publisher={APS}
}

@article{walshe2020continuous,
  title={Continuous-variable gate teleportation and bosonic-code error correction},
  author={Walshe, Blayney W and Baragiola, Ben Q and Alexander, Rafael N and Menicucci, Nicolas C},
  journal={Physical Review A},
  volume={102},
  number={6},
  pages={062411},
  year={2020},
  publisher={APS}
}

@article{kimble2008quantum,
  title={The quantum internet},
  author={Kimble, H Jeff},
  journal={Nature},
  volume={453},
  number={7198},
  pages={1023--1030},
  year={2008},
  publisher={Nature Publishing Group}
}

@article{wehner2018quantum,
  title={Quantum internet: A vision for the road ahead},
  author={Wehner, Stephanie and Elkouss, David and Hanson, Ronald},
  journal={Science},
  volume={362},
  number={6412},
  pages={eaam9288},
  year={2018},
  publisher={American Association for the Advancement of Science}
}

@article{pirandola2020advances,
  title={Advances in quantum cryptography},
  author={Pirandola, Stefano and Andersen, Ulrik L and Banchi, Leonardo and Berta, Mario and Bunandar, Darius and Colbeck, Roger and Englund, Dirk and Gehring, Tobias and Lupo, Cosmo and Ottaviani, Carlo and others},
  journal={Advances in optics and photonics},
  volume={12},
  number={4},
  pages={1012--1236},
  year={2020},
  publisher={Optical Society of America}
}

@article{scarani2009security,
  title={The security of practical quantum key distribution},
  author={Scarani, Valerio and Bechmann-Pasquinucci, Helle and Cerf, Nicolas J and Du{\v{s}}ek, Miloslav and L{\"u}tkenhaus, Norbert and Peev, Momtchil},
  journal={Reviews of modern physics},
  volume={81},
  number={3},
  pages={1301--1350},
  year={2009},
  publisher={APS}
}

@article{ma2012quantum,
  title={Quantum teleportation over 143 kilometres using active feed-forward},
  author={Ma, Xiao-Song and Herbst, Thomas and Scheidl, Thomas and Wang, Daqing and Kropatschek, Sebastian and Naylor, William and Wittmann, Bernhard and Mech, Alexandra and Kofler, Johannes and Anisimova, Elena and others},
  journal={Nature},
  volume={489},
  number={7415},
  pages={269--273},
  year={2012},
  publisher={Nature Publishing Group UK London}
}

@article{azuma2023quantum,
  title={Quantum repeaters: From quantum networks to the quantum internet},
  author={Azuma, Koji and Economou, Sophia E and Elkouss, David and Hilaire, Paul and Jiang, Liang and Lo, Hoi-Kwong and Tzitrin, Ilan},
  journal={Reviews of Modern Physics},
  volume={95},
  number={4},
  pages={045006},
  year={2023},
  publisher={APS}
}

@article{bhaskar2020experimental,
  title={Experimental demonstration of memory-enhanced quantum communication},
  author={Bhaskar, Mihir K and Riedinger, Ralf and Machielse, Bartholomeus and Levonian, David S and Nguyen, Christian T and Knall, Erik N and Park, Hongkun and Englund, Dirk and Lon{\v{c}}ar, Marko and Sukachev, Denis D and others},
  journal={Nature},
  volume={580},
  number={7801},
  pages={60--64},
  year={2020},
  publisher={Nature Publishing Group UK London}
}

@article{zhan2025experimental,
  title={Experimental demonstration of long distance quantum communication with independent heralded single photon sources},
  author={Zhan, Xiao-Hai and Zhong, Zhen-Qiu and Ma, Jian-Yu and Wang, Shuang and Yin, Zhen-Qiang and Chen, Wei and He, De-Yong and Guo, Guang-Can and Han, Zheng-Fu},
  journal={npj Quantum Information},
  volume={11},
  number={1},
  pages={73},
  year={2025},
  publisher={Nature Publishing Group UK London}
}

@article{azuma2015all,
  title={All-photonic quantum repeaters},
  author={Azuma, Koji and Tamaki, Kiyoshi and Lo, Hoi-Kwong},
  journal={Nature communications},
  volume={6},
  number={1},
  pages={6787},
  year={2015},
  publisher={Nature Publishing Group UK London}
}

@article{borregaard2020one,
  title={One-way quantum repeater based on near-deterministic photon-emitter interfaces},
  author={Borregaard, Johannes and Pichler, Hannes and Schr{\"o}der, Tim and Lukin, Mikhail D and Lodahl, Peter and S{\o}rensen, Anders S},
  journal={Physical Review X},
  volume={10},
  number={2},
  pages={021071},
  year={2020},
  publisher={APS}
}

@article{yamamoto1994quantum,
  title={The quantum optical repeater},
  author={Yamamoto, Yoshihisa},
  journal={Science},
  volume={263},
  number={5152},
  pages={1394--1395},
  year={1994},
  publisher={American Association for the Advancement of Science}
}

@article{dias2017quantum,
  title={Quantum repeaters using continuous-variable teleportation},
  author={Dias, Josephine and Ralph, Timothy C},
  journal={Physical Review A},
  volume={95},
  number={2},
  pages={022312},
  year={2017},
  publisher={APS}
}

@article{briegel1998interdisciplinary,
  title={Interdisciplinary Physics: Biological Physics, Quantum Information, etc-Quantum Repeaters: The Role of Imperfect Local Operations in Quantum Communication},
  author={Briegel, HJ and Dur, W and Cirac, JI and Zoller, P},
  journal={Physical Review Letters},
  volume={81},
  number={26},
  pages={5932--5935},
  year={1998},
  publisher={[Woodbury, NY, etc.] American Physical Society.}
}

@article{van2006hybrid,
  title={Hybrid quantum repeater using bright coherent light},
  author={Van Loock, P and Ladd, TD and Sanaka, K and Yamaguchi, F and Nemoto, Kae and Munro, WJ and Yamamoto, Y},
  journal={Physical review letters},
  volume={96},
  number={24},
  pages={240501},
  year={2006},
  publisher={APS}
}

@article{li2019experimental,
  title={Experimental quantum repeater without quantum memory},
  author={Li, Zheng-Da and Zhang, Rui and Yin, Xu-Fei and Liu, Li-Zheng and Hu, Yi and Fang, Yu-Qiang and Fei, Yue-Yang and Jiang, Xiao and Zhang, Jun and Li, Li and others},
  journal={Nature photonics},
  volume={13},
  number={9},
  pages={644--648},
  year={2019},
  publisher={Nature Publishing Group UK London}
}

@article{cochrane1999macroscopically,
  title={Macroscopically distinct quantum-superposition states as a bosonic code for amplitude damping},
  author={Cochrane, Paul T and Milburn, Gerard J and Munro, William J},
  journal={Physical Review A},
  volume={59},
  number={4},
  pages={2631},
  year={1999},
  publisher={APS}
}

@article{PhysRevA.64.012310,
  title = {Encoding a qubit in an oscillator},
  author = {Gottesman, Daniel and Kitaev, Alexei and Preskill, John},
  journal = {Phys. Rev. A},
  volume = {64},
  issue = {1},
  pages = {012310},
  numpages = {21},
  year = {2001},
  month = {Jun},
  publisher = {American Physical Society},
  doi = {10.1103/PhysRevA.64.012310},
  url = {https://link.aps.org/doi/10.1103/PhysRevA.64.012310}
}

@article{ofek2016extending,
  title={Extending the lifetime of a quantum bit with error correction in superconducting circuits},
  author={Ofek, Nissim and Petrenko, Andrei and Heeres, Reinier and Reinhold, Philip and Leghtas, Zaki and Vlastakis, Brian and Liu, Yehan and Frunzio, Luigi and Girvin, Steven M and Jiang, Liang and others},
  journal={Nature},
  volume={536},
  number={7617},
  pages={441--445},
  year={2016},
  publisher={Nature Publishing Group UK London}
}

@article{konno2024logical,
  title={Logical states for fault-tolerant quantum computation with propagating light},
  author={Konno, Shunya and Asavanant, Warit and Hanamura, Fumiya and Nagayoshi, Hironari and Fukui, Kosuke and Sakaguchi, Atsushi and Ide, Ryuhoh and China, Fumihiro and Yabuno, Masahiro and Miki, Shigehito and others},
  journal={Science},
  volume={383},
  number={6680},
  pages={289--293},
  year={2024},
  publisher={American Association for the Advancement of Science}
}

@article{fluhmann2019encoding,
  title={Encoding a qubit in a trapped-ion mechanical oscillator},
  author={Fl{\"u}hmann, Christa and Nguyen, Thanh Long and Marinelli, Matteo and Negnevitsky, Vlad and Mehta, Karan and Home, JP},
  journal={Nature},
  volume={566},
  number={7745},
  pages={513--517},
  year={2019},
  publisher={Nature Publishing Group UK London}
}

@article{fukui2021all,
  title={All-optical long-distance quantum communication with Gottesman-Kitaev-Preskill qubits},
  author={Fukui, Kosuke and Alexander, Rafael N and van Loock, Peter},
  journal={Physical Review Research},
  volume={3},
  number={3},
  pages={033118},
  year={2021},
  publisher={APS}
}

@article{rozpkedek2021quantum,
  title={Quantum repeaters based on concatenated bosonic and discrete-variable quantum codes},
  author={Rozp{k{e}}dek, Filip and Noh, Kyungjoo and Xu, Qian and Guha, Saikat and Jiang, Liang},
  journal={npj Quantum Information},
  volume={7},
  number={1},
  pages={102},
  year={2021},
  publisher={Nature Publishing Group UK London}
}

@article{schmidt2024error,
  title={Error-corrected quantum repeaters with Gottesman-Kitaev-Preskill qudits},
  author={Schmidt, Frank and Miller, Daniel and van Loock, Peter},
  journal={Physical Review A},
  volume={109},
  number={4},
  pages={042427},
  year={2024},
  publisher={APS}
}

@article{haussler2025quantum,
  title={Quantum repeaters based on stationary Gottesman-Kitaev-Preskill qubits},
  author={H{\"a}ussler, Stefan and van Loock, Peter},
  journal={Physical Review A},
  volume={111},
  number={6},
  pages={062611},
  year={2025},
  publisher={APS}
}

@article{ralph2024noise,
  title={Noise Transfer Approach to GKP Quantum Circuits},
  author={Ralph, Timothy C and Winnel, Matthew S and Swain, S Nibedita and Marshman, Ryan J},
  journal={Entropy},
  volume={26},
  number={10},
  pages={874},
  year={2024},
  publisher={MDPI}
}

@article{tzitrin2020progress,
  title={Progress towards practical qubit computation using approximate Gottesman-Kitaev-Preskill codes},
  author={Tzitrin, Ilan and Bourassa, J Eli and Menicucci, Nicolas C and Sabapathy, Krishna Kumar},
  journal={Physical Review A},
  volume={101},
  number={3},
  pages={032315},
  year={2020},
  publisher={APS}
}

@article{albert2018performance,
  title={Performance and structure of single-mode bosonic codes},
  author={Albert, Victor V and Noh, Kyungjoo and Duivenvoorden, Kasper and Young, Dylan J and Brierley, RT and Reinhold, Philip and Vuillot, Christophe and Li, Linshu and Shen, Chao and Girvin, Steven M and others},
  journal={Physical Review A},
  volume={97},
  number={3},
  pages={032346},
  year={2018},
  publisher={APS}
}

@article{fukui2017analog,
  title={Analog quantum error correction with encoding a qubit into an oscillator},
  author={Fukui, Kosuke and Tomita, Akihisa and Okamoto, Atsushi},
  journal={Physical review letters},
  volume={119},
  number={18},
  pages={180507},
  year={2017},
  publisher={APS}
}

@article{muralidharan2014ultrafast,
  title={Ultrafast and fault-tolerant quantum communication across long distances},
  author={Muralidharan, Sreraman and Kim, Jungsang and L{\"u}tkenhaus, Norbert and Lukin, Mikhail D and Jiang, Liang},
  journal={Physical review letters},
  volume={112},
  number={25},
  pages={250501},
  year={2014},
  publisher={APS}
}

@article{guanzon2022ideal,
  title={Ideal quantum teleamplification up to a selected energy cutoff using linear optics},
  author={Guanzon, Joshua J and Winnel, Matthew S and Lund, Austin P and Ralph, Timothy C},
  journal={Physical Review Letters},
  volume={128},
  number={16},
  pages={160501},
  year={2022},
  publisher={APS}
}

@article{swain2025enhancing,
  title={Enhancing Long-distance Continuous-variable Quantum-key-distribution with an Error-correcting Relay},
  author={Swain, S Nibedita and Marshman, Ryan J and Dias, Josephine and Solntsev, Alexander S and Ralph, Timothy C},
  journal={arXiv preprint arXiv:2512.11224},
  year={2025}
}

@article{munro2012quantum,
  title={Quantum communication without the necessity of quantum memories},
  author={Munro, William J and Stephens, Ashley M and Devitt, Simon J and Harrison, Keith A and Nemoto, Kae},
  journal={Nature Photonics},
  volume={6},
  number={11},
  pages={777--781},
  year={2012},
  publisher={Nature Publishing Group UK London}
}

@article{noh2020encoding,
  title={Encoding an oscillator into many oscillators},
  author={Noh, Kyungjoo and Girvin, SM and Jiang, Liang},
  journal={Physical Review Letters},
  volume={125},
  number={8},
  pages={080503},
  year={2020},
  publisher={APS}
}

@article{menicucci2014fault,
  title={Fault-tolerant measurement-based quantum computing with continuous-variable cluster states},
  author={Menicucci, Nicolas C},
  journal={Physical review letters},
  volume={112},
  number={12},
  pages={120504},
  year={2014},
  publisher={APS}
}

@article{noh2018quantum,
  title={Quantum capacity bounds of Gaussian thermal loss channels and achievable rates with Gottesman-Kitaev-Preskill codes},
  author={Noh, Kyungjoo and Albert, Victor V and Jiang, Liang},
  journal={IEEE Transactions on Information Theory},
  volume={65},
  number={4},
  pages={2563--2582},
  year={2018},
  publisher={IEEE}
}

@article{matsuura2020equivalence,
  title={Equivalence of approximate gottesman-kitaev-preskill codes},
  author={Matsuura, Takaya and Yamasaki, Hayata and Koashi, Masato},
  journal={Physical Review A},
  volume={102},
  number={3},
  pages={032408},
  year={2020},
  publisher={APS}
}

@article{larsen2021fault,
  title={Fault-tolerant continuous-variable measurement-based quantum computation architecture},
  author={Larsen, Mikkel V and Chamberland, Christopher and Noh, Kyungjoo and Neergaard-Nielsen, Jonas S and Andersen, Ulrik L},
  journal={Prx Quantum},
  volume={2},
  number={3},
  pages={030325},
  year={2021},
  publisher={APS}
}

@article{cohen2022low,
  title={Low-overhead fault-tolerant quantum computing using long-range connectivity},
  author={Cohen, Lawrence Z and Kim, Isaac H and Bartlett, Stephen D and Brown, Benjamin J},
  journal={Science Advances},
  volume={8},
  number={20},
  pages={eabn1717},
  year={2022},
  publisher={American Association for the Advancement of Science}
}

@article{duivenvoorden2017single,
  title={Single-mode displacement sensor},
  author={Duivenvoorden, Kasper and Terhal, Barbara M and Weigand, Daniel},
  journal={Physical Review A},
  volume={95},
  number={1},
  pages={012305},
  year={2017},
  publisher={APS}
}

@article{varnava2006loss,
  title={Loss tolerance in one-way quantum computation via counterfactual error correction},
  author={Varnava, Michael and Browne, Daniel E and Rudolph, Terry},
  journal={Physical review letters},
  volume={97},
  number={12},
  pages={120501},
  year={2006},
  publisher={APS}
}

@article{fukui2023efficient,
  title={Efficient concatenated bosonic code for additive Gaussian noise},
  author={Fukui, Kosuke and Matsuura, Takaya and Menicucci, Nicolas C},
  journal={Physical Review Letters},
  volume={131},
  number={17},
  pages={170603},
  year={2023},
  publisher={APS}
}

@article{braunstein2005quantum,
  title={Quantum information with continuous variables},
  author={Braunstein, Samuel L and Van Loock, Peter},
  journal={Reviews of modern physics},
  volume={77},
  number={2},
  pages={513--577},
  year={2005},
  publisher={APS}
}

@article{slussarenko2022quantum,
  title={Quantum channel correction outperforming direct transmission},
  author={Slussarenko, Sergei and Weston, Morgan M and Shalm, Lynden K and Verma, Varun B and Nam, Sae-Woo and Kocsis, Sacha and Ralph, Timothy C and Pryde, Geoff J},
  journal={Nature Communications},
  volume={13},
  number={1},
  pages={1832},
  year={2022},
  publisher={Nature Publishing Group UK London}
}

@article{lund2008fault,
  title={Fault-tolerant linear optical quantum computing with small-amplitude coherent states},
  author={Lund, Austin P and Ralph, Timothy C and Haselgrove, Henry L},
  journal={Physical review letters},
  volume={100},
  number={3},
  pages={030503},
  year={2008},
  publisher={APS}
}

@article{glancy2006error,
  title={Error analysis for encoding a qubit in an oscillator},
  author={Glancy, Scott and Knill, Emanuel},
  journal={Physical Review A—Atomic, Molecular, and Optical Physics},
  volume={73},
  number={1},
  pages={012325},
  year={2006},
  publisher={APS}
}

@misc{winnel2021,
      title={Overcoming the repeaterless bound in continuous-variable quantum communication without quantum memories}, 
      author={Matthew S. Winnel and Joshua J. Guanzon and Nedasadat Hosseinidehaj and Timothy C. Ralph},
      year={2021},
      eprint={2105.03586},
      archivePrefix={arXiv},
      primaryClass={quant-ph},
      url={https://arxiv.org/abs/2105.03586}, 
}

\end{document}